\documentclass[11pt, a4paper, reqno]{article}
\usepackage{amsmath, amsfonts, amsthm, amssymb, here, dsfont, hyperref}
\usepackage{graphicx,color,multirow, here, subcaption,caption}
\usepackage{mathtools}
\usepackage[utf8]{inputenc} 
\usepackage[T1]{fontenc}
\usepackage{hhline}
\usepackage[english]{babel}
\usepackage{here,dsfont, hyperref}
\usepackage{natbib}
\usepackage{url}
\usepackage{bbm}
\usepackage{appendix}
\usepackage{array}
\usepackage[colorinlistoftodos,textwidth=2.3cm]{todonotes}

\newcommand{\argmax}[1]{\underset{#1}{\operatorname{arg}\!\operatorname{max}}\;}

\newcolumntype{C}[1]{>{\centering\let\newline\\\arraybackslash\hspace{0pt}}m{#1}}

\newcommand{\w}{\widehat}

\newcommand{\E}{\mathbb{E}}
\newcommand{\cE}{\mathcal{E}}

\renewcommand{\P}{\mathbb{P}}

\newcommand{\Nt}{N(t)}
\newcommand{\phigamma}{\phi_{\gammaij }}
\newcommand{\aij}{a_{ij}}
\newcommand{\bij}{b_{ij}}
\newcommand{\gammaij}{\gamma_{ij}}
\newcommand{\lambdatheta}[1]{\lambda_{\theta}^{#1}}
\newcommand{\Lambdatheta}[1]{\Lambda_{\theta}^{#1}}

\newcommand{\Ni}{N^{i}}
\newcommand{\Nj}{N^{j}}
\newcommand{\Tk}{T_{k}}
\newcommand{\Tik}{T_{k}^i}
\newcommand{\indic}{\ensuremath{\mathds{1}}}
\newcommand{\setindex}{\in \{1,.., d\}}

\newcommand{\Hz}{\ensuremath{\mathrm{H_0}}}
\newcommand{\Hu}{\ensuremath{\mathrm{H_1}}}

\newtheorem{theo}{Theorem}[section]
\newtheorem{defi}[theo]{Definition}
\newtheorem{prop}[theo]{Proposition}

\newtheorem{rem}[theo]{Remark}

\newtheorem{ass}[theo]{Assumption}

\usepackage{thmtools}
\declaretheorem[
  shaded={rulecolor=black, rulewidth=1pt, bgcolor=white},
  name=Test]{test}

\def\limiteloi{\renewcommand{\arraystretch}{0.5}
\begin{array}[t]{c}\stackrel{{\cal L}}{\longrightarrow} \\
{\scriptstyle T\rightarrow+\infty}\end{array}\renewcommand{\arraystretch}{1}}
\def\limiteproba{\renewcommand{\arraystretch}{0.5}
\begin{array}[t]{c}\stackrel{{\P}}{\longrightarrow} \\
{\scriptstyle T\rightarrow+\infty}\end{array}\renewcommand{\arraystretch}{1}}

\renewcommand{\P}{\mathbb{P}}

\usepackage{comment}

\title{Testing procedures based on maximum likelihood estimation for Marked Hawkes processes}
\author{Anna Bonnet$^{(1)}$, Charlotte Dion-Blanc$^{(1)}$, Maya Sadeler Perrin$^{(2)}$
\vspace{2mm}\\
(1) LPSM, UMR 8001, Sorbonne Universit\'e, 75005 Paris, France \\
(2) LJK, UMR 5224, Univ. Grenoble Alpes, Grenoble INP, 38000 Grenoble, France
}
\date{\today}

\begin{document}

\maketitle

\begin{abstract}
    The Hawkes model is a past-dependent point process, widely used in various fields for modeling temporal clustering of events. Extending this framework, the multidimensional marked Hawkes process incorporates multiple interacting event types and additional marks, enhancing its capability to model complex dependencies in multivariate time series data. However, increasing the complexity of the model also increases the computational cost of the associated estimation methods and may induce an overfitting of the model. Therefore, it is essential to find a trade-off between accuracy and artificial complexity of the model. In order to find the appropriate version of Hawkes processes, we address, in this paper, the tasks of model fit evaluation and parameter testing for marked Hawkes processes. This article focuses on parametric Hawkes processes with exponential memory kernels, a popular variant  for its theoretical and practical advantages.  Our work introduces robust testing methodologies for assessing model parameters and complexity, building upon and extending previous theoretical frameworks. We then validate the practical robustness of these tests through comprehensive numerical studies, especially in scenarios where theoretical guarantees remains incomplete.
\end{abstract}



\section{Introduction}


\paragraph{State of the art.}
The Hawkes process, introduced by \cite{hawkes1971point}, is a self-exciting point process, where each event increases the likelihood of future events occurring. It is widely used in various fields due to its ability to model clustering of events over time. Unlike Poisson processes, Hawkes processes allow for the inclusion of past events' influence on future occurrences, providing a more realistic framework for temporal event data. A natural extension of the univariate Hawkes process is the multidimensional Hawkes process \citep{ogata1988statistical}, which models interactions among multiple types of events. In this framework, events in one dimension can excite or inhibit events in other dimensions, capturing complex dependencies and interactions in multivariate time series data \citep{carstensen2010multivariate, bonnet2023inference,sulem2024bayesian}. 

The multidimensional marked Hawkes process extends the multidimensional Hawkes model to multiple interacting point processes with additional marks that carry extra information about each event \citep{zhuang2002stochastic}. Each dimension represents a different type of event, and the occurrence of an event in one dimension can increase or decrease the probability of events in other dimensions, with a strength that depends on the value of the mark \citep{reinhart2018review}. These marks, which can include features such as event size or type, provide further information to the model. Such a framework is particularly useful in applications like financial markets \citep{lotz2024sparsity} and seismology \citep{zhuang2002stochastic}, where external variables can significantly impact the probability of future events.

As Hawkes processes are characterized by their intensity function, which describes the infinitesimal probability of a new event occurring, conditionally on the past, estimating these processes boils down to estimate this function. A wide range of techniques exists, depending on the model's flexibility and on the available data. In the parametric version of the process, the standard approach relies on the maximum likelihood method to estimate the parameters \citep{ogata1978estimators, bonnet2023inference}. Other methods, such as the EM algorithm \citep{veen2008estimation} or least squares in \cite{bacry2015sparse}, are also used, though they are mainly applied when the process is unmarked. On the other hand, when the intensity is defined nonparametrically, common methods include non-parametric EM algorithms \citep{lewis2011nonparametric}, Bayesian inference \citep{deutsch2022bayesian, sulem2024bayesian}, or RKHS techniques \citep{yang2017online}, which have been widely studied, especially in unmarked cases. 

In this article, we focus on the specific case of parametric Hawkes processes, where the intensity function has an exponential memory kernel. This variant of the Hawkes process is widely used, see \emph{e.g.} \cite{zhang2020self}, as it satisfies the Markov property, thanks to the simple factorization of its exponential function, which is beneficial both in theory and  practice.
Even within this subset of Hawkes processes, there is significant model diversity: exciting, inhibiting, or marked versions of the process. Given this plurality, a crucial yet often underestimated step in the analysis of these processes is evaluating the quality of model fit and the accuracy of the estimated coefficients. As many of the considered models are nested, each increase in complexity introduces additional parameters, leading to greater computational demands and potential overfitting. Therefore, developing robust testing and selection procedures is essential.

Currently, some test procedures exist in the literature, one can cite for example \cite{daley2008introduction,reynaud2014goodness, clinet2021asymptotic}. Recently, in \cite{lotz2024sparsity}, the author developed likelihood-ratio tests to evaluate the sparsity of the coefficients matrix that describes how subprocesses interact with each other. This work focuses on the case where only self-exciting phenomena are possible, and where repeated observations of the same process are available. Moreover, main references \citep{clinet2021asymptotic,richards2022score} operate in a framework where the temporal dynamics of the process are partially known or lack numerical studies to evaluate their performance, as they are typically examined in a highly theoretical context. As a result, some commonly used tests for assessing the quality of fit have been shown to exhibit biases in their typical applications \citep{reynaud2014goodness}. Finally, recently, tests based on bootsrap procedures have been proposed in \cite{cavaliere2023bootstrap}. 

\paragraph{Our contribution.} 

 
With this article, we aim at proposing a unified framework to perform tests procedures on Hawkes processes. Due to the abundance of inference procedures proposed in many different frameworks - Bayesian, frequentist, parametric, nonparametric - we chose to restrict our analysis to procedures based on maximum likelihood approaches and its derivatives. First, we propose a full pipeline with a corresponding numerical implementation that contains many extensions of the existing literature that either does not provide a code or only for limited scenarios. For instance, standard procedures frequently require the computation of the compensator or the process, which can be highly non-trivial for complex models, e.g. the nonlinear marked Hawkes process. Our implementation also integrates state-of-the-art techniques such as the subsampling method \citep{reynaud2014goodness} or the \texttt{qqconf} confidence intervals \citep{weine2023application}. 

In addition, we want to underline the restricted theoretical setting in which most of the existing procedures have been developed, extend it when possible and exhibit its limitations otherwise. 

Finally, we want to assess the empirical performance of the considered methodologies outside their theoretical setting in order to highlight for potential misuse or to motivate the development of further theoretical grounds. In practice, we distinguish between two types of tests. The first one requires only a single observation and gather procedures that are derived from the asymptotic properties of maximum likelihood estimators. The second one, usually called goodness-of-fit procedures since they allow to assess the fit between a proposed model and the data, can generalize to many models and inference methods but requires several observations of the process, which are not always available in practice.


\paragraph{Organization of the article.} 

The article is organized as follows. In Section \ref{Section:Model}, we present the different models under study, while Section \ref{Section:TheoreticalProperty} gahters theoretical results for estimators and tests based on the likelihood, for the different models presented in Section \ref{Section:Model}. In Section \ref{Section:Test_Procedure}, we present the test procedures associated with each model. In Section \ref{Section:Simulation} we conduct a numerical study of the presented test procedures on synthetic data. Python's codes developed for those simulations are freely available on the git repository at the address: \url{https://github.com/Msadeler/marked_exp_hawkes}. Finally, we illustrate several testing procedures on two different data sets: a seismology database and a neural database.

\section{Exponential Hawkes Processes}
\label{Section:Model}

In this section, we present the different models considered in the following. 
These models, which are derived from Hawkes processes, are in fact part of a wider range of objects known as point processes.  To introduce Hawkes processes, we first introduce some concepts and notation associated with the theory of point processes. For a fuller understanding of this theory and these objects, the reader is referred to \cite{daley2003introduction} and \cite{daley2008introduction}.

\paragraph{Notation for point process.} The Borel algebra is denoted
$\mathcal{B}(\mathbb{R}_+)$. The point process $N=(\Nt)_{t\geq 0}$ is of dimension $d \geq 1$, and its components are denoted $(N^i)_{1\leq i \leq d}$. The associate filtration is $(\mathcal{F}_t):= \sigma( N(s), ~ s < t)$. 
 In the broader context of point processes, the characterization of these processes relies on the definition of their conditional intensity function denoted $t \mapsto \lambda(t)$, which is $\mathcal{F}_t$ predictable, and defined as
\[
\lambda(t) = \lim_{{h \rightarrow 0^+}} \frac{1}{h} \mathbb{E}\left [N(t+h) - N(t) \, \vert \, \mathcal{F}_{t^-}\right].
\]
The associate compensator is defined as
\begin{equation}
\label{eq:Compensator}
    \Lambda(t) = \int_{(0,t) }\lambda(s)ds, 
\end{equation}
with a theoretical definition that can be found in \cite{meyer1962decomposition}. Process  $(\Lambda(t))_{t \geq 0}$ is a right continuous function such that $(\Nt - \Lambda(t))_{t>0}$ is a local martingale. 
We denote $(\kappa_i)$ the external variables, which law does not depend on the process $N$ but can influence it. Those variables live in a separable complete metric space denoted $\cE$. We adopt a frequentist approach in the remainder of the article. Consequently each model presented depends on a parameter that is denoted $\theta$.  This parameter lies down in a set $\Theta$ that, given the model under consideration, can take the form  $ (\mathbb{R}_+^* \times \mathbb{K}^d \times (\mathbb{R}_+^*)^d \times \cE^d)^d$ or $(\mathbb{R}_+^* \times \mathbb{K}^d \times (\mathbb{R}_+^*)^d)^d$, with $\mathbb{K}$ being $\mathbb{R}$ or $\mathbb{R}_+$.  The intensity and compensator are denoted $\lambdatheta{}$ and $\Lambdatheta{}$, respectively.

In this section, we introduce the Linear Multidimensional Marked Exponential Hawkes Process (Linear Multidimensional Marked EHP), which constitutes the object of investigation of this paper. To state its conceptual framework and utility, we first present a simpler, well-established variant, known as the Linear Multidimensional Exponential Hawkes Process (Linear Multidimensional EHP).

\paragraph{Linear Multidimensional  EHP.}
We denote $\theta_i = (m_i, (\aij)_{j \setindex }, (\bij)_{j \setindex })$  with, for all $i,j \setindex $ $m_i \geq 0$, $\bij  > 0$, and $\aij \geq 0$ and $\theta = (\theta_1, ...., \theta_d) \in \Theta$. A Linear Multidimensional Exponential Hawkes Process is defined as a vector of $d$ point processes, denoted by $\Nt = (N^1(t), \ldots , N^d(t))$, each component having a conditional intensity $\lambdatheta{i}$, given for $t>0$ by the following equation,
\begin{equation}
\label{eq:LMEHP}
\lambdatheta{i}(t) =  
 m_i +  \sum_{j=1}^{d} \int_{(-\infty,t)} \aij e^{-\bij ( t-s )} \Nj(ds). 
\end{equation}

This process can either be represented by the vector $\Nt$ or, by the sequence of events $(\Tk, m_k)_{k \in \mathbb{N}}  \in (\mathbb{R} \times \mathbb{N})^{\mathbb{N}}$, where $(\Tk)$ are the ordered arrival times and $m_k$ represents the associated component with each arrival, as $\Ni(t)= \sum_{m_k = i }  \mathbbm{1}_{T_k< t }$. Throughout this article, we interchangeably use the notations $N$ or $(T_{i}, m_i)$ to describe a point process. We also sometimes denote $(\Tik)_k$ the events associated with the $i-th$ component of $\Nt$. 

The expression of the intensity function underscores the strong temporal interdependencies driving arrival times, as the probability of a new arrival in component $i$ during $[t, t+dt]$ is approximately determined by $\lambdatheta{i}(t)dt$. Consequently, each past arrival influences the probability of future events. In Equation \eqref{eq:LMEHP}, three parameters have an influence on $\lambdatheta{i}$ (and thereby on the arrival probability), each carrying distinct significance: $m_i$ represents the base excitation rate of the process $N^i$, while $\aij$ denotes the strength of the interaction between processes $N^i$ and $\Nj$, which duration is regulated by $\bij $.

However, such models cannot accommodate external variables, as each sub-process can only be influenced by others. To address this limitation, we introduce the Linear Multidimensional Marked EHP which is defined as follows.

\paragraph{Linear Multidimensional Marked EHP.}

\begin{defi} [Linear Multidimensional Marked Exponential Hawkes Process]\label{def:LMMEHP} We denote $\theta_i = (m_i, (\aij)_{j \setindex }, (\bij)_{j \setindex })$  with, for all $i,j \setindex $ $m_i \geq 0$, $\bij  > 0$, and $\aij \geq 0$ and $\theta = (\theta_1, ...., \theta_d) \in \Theta$. Consider a $d$-dimensional punctual process $\Nt=(N^1(t), ..., N^d(t))$ associated with the event $(T_i, m_i, \kappa_i)$ where $T_i$ is the event time, $m_i$ the associated component and $\kappa_i$ the associated mark.
This process is a Linear Marked Multidimensional  Exponential Hawkes Process if, for all $i \setindex $ the conditional intensity of the $i$-th sub-process can be written
\begin{equation}
\label{defLamba_mark}
   \lambdatheta{i}(t) =   m_i + \sum_{j=1}^{d} \int_{(-\infty ,t)\times \cE} \aij e^{-\bij(t-s)} \phi_{ij}(\kappa) \Nj(ds \times d\kappa),
\end{equation}
with $\phi_{ij}: \cE \rightarrow  \mathbb{R}_+^*$.
\end{defi}
In the rest of the article, we only consider marks that are i.i.d. random variables, denoted $(\kappa_i)$.
Definition \eqref{defLamba_mark} introduces a more intricate expression of process intensity, providing a richer framework as the random variables $(\kappa_i)$, namely the marks, can influence the intensity of interaction between sub-processes: the coefficient $\aij$ is modulated by the value $\phi_{ij}(\kappa_i)$, allowing the mark to attenuate ($\phi_{ij}(\kappa_i) < 1$) or amplify ($\phi_{ij}(\kappa_i) > 1$) the interaction.

In the parametric formulation of the marked model, the functions $\phi_{ij}$ are selected based on a parameter $\gammaij  \in \mathbb{R}$ and we thus denote $\phi_{ij}:= \phigamma$ and $\theta_i = ( m_i, (\aij)_{j \setindex}, (\bij)_{j \setindex}, (\gammaij)_{j \setindex} )$. Common cases include $\phigamma$ to be equal to $ x \mapsto e^{\gammaij x}$ or $ x \mapsto x^{\gammaij }$. To ensure that this model is nested in the Linear Multidimensional Marked EHP model, we suppose that,  for $\gamma =0$, the function $\phi_{\gamma}$ is the constant function equal to one.

With those assumptions, all models displayed in this section are nested: the marked model is an extension of the unmarked one, which is itself an extension of the classical Poisson model. As we formulate those models, we first demonstrate identifiability in the linear cases.

\begin{prop}
    \label{Proposition:identifiabilite}
    In a Linear Multidimensional Marked EHP, let $\phi_{\gammaij }$ be, for all $i,j \setindex$ of the form $x \mapsto e^{\gammaij  x}$ or $x \mapsto x^{\gammaij }$. If, for all $i \in {1,...,d}$, for all $j \neq i$, there exist $l,l' >0 $ such that $T_{\ell}$ and $T_{l'}$ are arrival times of component $j$ satisfying $\kappa_l \neq \kappa_{l'}$ and $\kappa_l, \kappa_{l'} >0$, then the model is identifiable.
\end{prop}
The proof of this result is relegated in Appendix. In this proof, it is trivial that the coefficients $m_{i}$ and $\bij$ are identifiable as soon as the concerned component has at least one jump time. The second condition serves to identify $\gammaij $ and $\aij$. Thus, the condition for having an identifiable model is to have enough jumps for each component, but also to have sufficiently different marks at each arrival time to reconstruct each function $\phi_{\gammaij }$. Typically, if we take marks that are i.i.d. and have density with respect to Lebesgue measure, the condition is almost surely satisfied as long as all each component have at least two jumps.
\begin{rem}
Here, we only need to have two different values for the mark, as the functions $\phi$ are uniquely determined by their evaluation at one point. If we were to take more complex forms for the function $\Phi: \gamma \mapsto \phi_\gamma$, we could still have model identifiability provided that $\Phi: \gamma \in \cE \mapsto \phi_\gamma \in \mathcal{C}(K, \mathbb{R}_+^*)$ is a one-to-one morphism from $(E, +)$ into $( \mathcal{C}(K, \mathbb{R}_+^*), \times) $, that all $\phi_\gamma$ are unitary and that the function $\kappa \mapsto \phi_0(\kappa)$ is determined by a finite number of its evaluation at distinct points.
\end{rem}

Lastly, we outline conditions to guarantee that a stationary version of the process exists and is observed if the observation time is large enough. This assumption, as detailed in \cite{embrechts2011multivariate}, imposes constraints on the ratio $\frac{\aij}{\bij }$, which is the usual condition to ensure stationarity of a Linear Multidimensionnal EHP, and the average value of the random variable $\phi_\gamma(\kappa)$, depending on the density $f$ of the marks according to a reference measure on $\cE$.

\begin{ass}
\label{Assumption:stationnarity}
    For all $i,j \setindex$, $\mathbb{E}_f \left[  \phigamma ( \kappa)\right]=1$ and the spectral radius of $Q = (\aij/\bij )_{ij}$ is strictly less than one.
\end{ass}

Once again, using a parametric framework, we suppose that the density is parameterized by $\psi \in \mathbb{R}$. As a result, Assumption \ref{Assumption:stationnarity} implies that the function $\phi$ depends on both parameters $\gammaij $ and $\psi$.
When we work under this assumption, we write 
$$\phi_{ij}:= \phi_{\gammaij , \psi}.$$

It is important to note that the introduction of this condition into our model prevents us from demonstrating its identifiability without further assumptions on the type of density for $(\kappa_i)$. This comes from the normalization condition on $\phi_\gamma$ against the mark density, which consequently introduces a normalization coefficient linked to the parameter defining the mark, $\psi$, and to $\gammaij$. Without further assumptions, it is challenging to distinguish the two coefficients, since the process intensity only gives access to $\aij \phi_{\gamma, \psi}(\kappa_i)$  In addition, this condition greatly complicates the forms of the functions $\phi_{\gamma, \psi}$, as this assumption as the normalization constant can be complex to compute explicitly, depending on the chosen density type.

\paragraph{Non-Linear Marked Multidimensional  EHP.}
We present a final version of a Hawkes Process, which extends the previous models by allowing the process to exhibit inhibition through $a_{ij}$ parameters that can be negative. Although there is no theoretical analysis of this model, it is of great practical interest for modeling phenomena with inhibition. In this paper, it will only be addressed from a numerical perspective, in Section \ref{Section:Simulation}.

\begin{defi} [Non-Linear Multidimensional Marked Exponential Hawkes Process]\label{def:NLMMEHP} 
We denote $\theta_i = (m_i, (\aij)_{j \setindex }, (\bij)_{j \setindex })$  with, for all $i,j \setindex $ $m_i \geq 0$, $\bij  > 0$, and $\aij \in \mathbb{R}$ and $\theta = (\theta_1, ...., \theta_d) \in \Theta$. Consider a $d$-dimensional point process $\Nt=(N^1(t), ..., N^d(t))$ associated with the event $(T_i, m_i, \kappa_i)$ where $T_i$ is the event time, $m_i$ the associated component and $\kappa_i$ the associated mark.

This process is a Non-Linear Marked Multidimensional  Exponential Hawkes Process (Linear Marked Multidimensional  EHP) if, for all $i \setindex $ the conditional intensity of the $i$-th sub-process can be written
\begin{equation}
\label{defLamba_markNL}
   \lambdatheta{i}(t) = \max \left( \lambdatheta{i, \star}(t), 0 \right) ~ \text{  with } ~ \lambdatheta{i, \star}(t) = m_i + \sum_{j=1}^{d} \int_{(-\infty ,t)\times \cE} \aij e^{-\bij(t-s)} \phi_{ij}(\kappa) \Nj(ds \times d\kappa)
\end{equation}
and $\phi_{ij}: \cE \rightarrow  \mathbb{R}_+^*$.
\end{defi}
If the stationary property of this process can be adapted from Assumption \ref{Assumption:stationnarity} by replacing $a_{ij}$ with $\vert a_{ij} \vert$, the identifiability of the model still needs to be proven, even without considering Assumption \ref{Assumption:stationnarity}. Identifiability has only been proven for the Non-Linear EHP (specifically when $\gamma_{ij} = 0$) as shown by \cite{bonnet2023inference}, under certain assumptions on the process.

\section{Maximum Likelihood Estimator and its theoretical properties}\label{Section:TheoreticalProperty}

In this Section, we compile a collection of theoretical findings for the Multidimensional Hawkes Process, whether marked or unmarked. In addition to presenting previously known theorems, we introduce some extensions tailored to the marked processes. Using this approach, we aim to provide a comprehensive and cohesive overview in a field where the literature is often fragmented. Those theoretical properties all play a role in estimating the parameters of models defined by Equations \eqref{eq:LMEHP} and \eqref{defLamba_mark}. Given that \eqref{eq:LMEHP} is a submodel of \eqref{defLamba_mark}, most theoretical properties are stated for \eqref{defLamba_mark} and thus apply to both models, unless specified otherwise.

\subsection{Maximum Likelihood Estimator}

We adopt a frequentist approach, assuming the existence of a true parameter $\theta^*$ in $\Theta$, such that the studied process has an intensity of the form $\lambda_{\theta^*}$ that we aim at estimating. In this context, the log-likelihood has a well known expression that is a direct function of the intensity and the density of the mark \citep{liniger2009multivariate}.
\begin{theo}[Log-likelihood of a Non-Linear Multidimensional Marked EHP]
\label{theo:LikelihoodMark}
 Let  $N$ be a Non-Linear Multidimensional Marked EHP and $\{ (\Tk, m_k, \kappa_k) \}$ the times of arrival with associated  component and mark, observed on $[0,T]$. Let $\theta \in \Theta$, the log-likelihood of the model defined by $\theta$ is given by,
 \begin{equation}
    \label{eq:Liklihood-Mark-Multi}
\ell_T( \theta, \psi):= \sum_{j=1}^{d} \left( \int_{0}^T \int_{\cE} \log(\lambdatheta{j}(s)) \Nj(ds \times d\kappa)  +  \int_{0}^T \int_{\cE} \log(f_\psi(\kappa)) \Nj(ds \times d\kappa) -  \int_{0}^T\lambdatheta{j}(s)ds \right).
\end{equation}

\end{theo}

The log-likelihood presented here relies on what is called the exact intensity, which is the intensity defined using the finite past framework (as the integral no longer starts at $t=-\infty$ but at $t=0$). We refer to \cite{ogata1978estimators} for the fact that considering the likelihood with the information from the infinite past (in the intensity process) is equivalent with the present definition, as long as we consider a stationary process. We thus are led to consider the maximum likelihood estimator that is the estimator maximizing the function $\ell_T$: \begin{equation}
    (\w{\theta}_T, \w{\psi}_T) \in \argmax{\theta,\psi \in \Theta \times \mathbb{R} } \ell_T(\theta, \psi).
\end{equation}
The likelihood decomposes into two terms, one term representing the density of the mark denoted $\ell^f_T(\psi)$, and one term representing the likelihood of the point process denoted $\ell^{N}(\theta)$:
$$ \ell_T(\theta, \psi) = \ell^{N}(\theta,\psi) + \ell^f_T(\psi).$$
As a result, when Assumption \ref{Assumption:stationnarity} is not respected and $\phi$ does not depend on $\psi$ it is possible to separate the estimation of the parameter $\theta$ from the estimation of $\psi$.  If the law of the mark is not of interest, which is our case as soon as this Assumption is violated, we omit the likelihood of the mark and do not estimate it.

\subsection{Consistency and asymptotic normality}

Regarding the literature on the MLE estimator, various properties can be found, but we focus here on the consistency and asymptotic behavior of the estimator, the first paper on the subject being \cite{ogata1978estimators}. However, the convergence conditions established by \cite{ogata1978estimators} are challenging to verify. In \cite{clinet2021asymptotic}, more accessible hypotheses are proposed, and it is shown that the multidimensional linear Hawkes process satisfies these conditions Yet, the assymptotic normal behavior of the MLE has not been extended to the Linear Multidimensional Marked EHP case. Therefore, the property of asymptotic normality is stated only for the linear unmarked model, and we refer to \cite{clinet2021asymptotic} for a demonstration of the following theorem.

\begin{theo} 
\label{Theorem:ConsistanceMLE}
Consider a Linear Multidimensional EHP and suppose that the Fisher information matrix, defined by $$\Gamma_{\theta^*}:= -\E_{\theta^*}\left[\partial_{\theta \theta}^2 \ell\left(\theta^*\right)\right]=\E_{\theta^*}\left[\int_{[0, T]} \lambda_{\theta^*}\left(t \right)^{-2}\left(\partial_\theta \lambda_{\theta^*}(t )\right)^{\otimes 2} N(d t)\right],$$ is not singular, where $\theta^\star$ is the true parameter defining the process.
Then, the MLE is consistent:
$$\w{\theta}_T \limiteproba  \theta^\star.$$
Furthermore:
\begin{equation}
    \label{CvLoiNormal}
        \sqrt T (\w{\theta}_T- \theta^*) \limiteloi N(0,\Gamma^{-1}_{\theta^*}).
    \end{equation}
And, the following convergence holds:        
\begin{equation}
    \label{CVHessianTowardFicher}
        \left(-\frac{1}{T} \partial_\theta^2 \ell_T(\w{\theta}_T)\right)^{-1} \limiteproba \Gamma_{\theta^*}^{-1} ,
    \end{equation}
where $\partial^2_{\theta}\ell_{T}(\w{\theta}_T)$ is the Hessian matrix of $\ell_{T}(\theta)$, w.r.t. $\theta$ evaluated in $\w{\theta}_T$.
\end{theo}  

For the rest of the paper, we denote $\w{I}$ any consistent approximation of the Fisher information matrix. Following the previous proposition, $\w{I}$ often refers to:
\begin{equation}
    \label{eq:Ihat}
    \w{I}:= -\frac{1}{T} \partial_\theta^2 \ell_T(\w{\theta}_T),
\end{equation}
but other consistent approximations exist, as $$\w{I}:= \frac{1}{T} \int_0^T \lambda_{\w{\theta}}(t)^{-2} \partial_\theta \lambda_{\w{\theta}}(t) \partial_\theta \lambda_{\w{\theta}}(t)^T N(dt)$$ that is used in \cite{richards2022score}.
Theorem \ref{Theorem:ConsistanceMLE}, though not explicitly featured in \cite{clinet2021asymptotic}, consolidates two results from the article. While \cite{clinet2021asymptotic} did not explicitly state this theorem, this specific formulation allows the construction of tests and confidence intervals discussed in subsequent sections.

It should be noted that even though $T$ appears in the definition of the Fisher matrix used, this dependence is artificial due to the assumption of stationarity.

\subsection{Rescaling: the time-change theorem}

Another major result for point processes is the \textit{time-change} theorem. It allows, using the compensator of the model, defined by Equation \eqref{eq:Compensator}, to transform the initial point process into a unit rate Poisson process. As we see later, this is a very important property that allows to assess the quality of the fitted model.

Here, we present a time-change model associated with the Linear Multidimensional Marked EHP model, that is a slightly different version from the time-change theorem that was originally stated in \cite{daley2008introduction}, yet the proof follows very similar steps. To extend this result to the non-linear model, it would be necessary to demonstrate that the process compensator tends towards plus infinity, which is not guaranteed in cases where the inhibition would be very strong relative to the base emission rate.

\begin{theo}
\label{Theorem:TimeChangeMark}
Let $(T_j, m_j, \kappa_j)_j$ be Linear Multidimensional Marked EHP process with $(\kappa_j) $ taking values in $\mathbb{R}$. 
We define $\Lambdatheta{i}(t): =  \int_{ (0,t)} ~ \lambdatheta{i}(s) ds$, $f$ the density function associated with the mark with respect to the Lebesgue measure, and $F$ as the corresponding distribution function. 

Suppose $\kappa \mapsto F(\kappa) $ is $\mathcal{C}^0$, and that, for all $i$, $t \mapsto \Lambdatheta{i}(t)$ is continuous and tends to infinity at infinity. Then, for any strictly increasing sequence $T_j^i$ in $\mathbb{R}_+^*$, $\Lambdatheta{i}(T_j^i)$ is a realization of a unit-rate Poisson process if and only if $(T_j^i)$ is a realization of the point process defined by $\lambdatheta{i}$.

Similarly, if we define $\widetilde{N}(t)= \sum_{i=1}^{d} \Ni(t),  ~ \Lambda_{\theta}: t \mapsto \sum_{i=1}^{d} \Lambdatheta{i}$, and assume this function is continuous, increasing, and tends to infinity at infinity, then $(T_j)$ is a realization of the process $\widetilde{N}(t)$ if and only if $\overline{N}:= (\Lambdatheta{}(T_j), F ( \kappa_j))$ is a realization of a unit-rate Poisson process.
\end{theo}

The proof is relegated in Appendix \ref{appendix:proof_time_change}.



\section{Test procedures}
\label{Section:Test_Procedure}

Building upon the theoretical results presented in the previous section, we present in the following several testing procedures to evaluate the Hawkes model validity. This framework encompasses both coefficients testing and model evaluation, particularly when considering the various nested models within the Hawkes framework, such as the Poisson model, the linear model, and the marked model. Our objective is to establish a clear methodology for conducting these tests, while justifying, when applicable, of the associated theoretical properties. Although the underlying theoretical foundations of these tests are not necessarily novel, many procedures have either not been formulated as tests, or have only been proposed in the simplest versions of Hawkes processes, typically univariate, linear, unmarked. Here, we aim at referencing various procedures, presenting the existing versions in the literature and the proposed extensions and validations.
    We also combine testing procedures and state-of-the-art statistical tools to enhance their performance.
For all the tests presented here, and for the rest of the article, we denote $\alpha$ the level of the test.

\subsection{Test relying on a single observation of the process}

When only one observation of the process is available, an intuitive way to conduct a test is to rely on the confidence intervals described in the previous section, based on the convergence in infinite time of the estimator. Given the different properties outlined in Section \ref{Section:TheoreticalProperty}, we are able to construct tests in the context of long time periods, where convergence to normality is nearly achieved. 

\subsubsection{Tests to evaluate and compare coefficient values}
\label{subsection:test_on_one_coeff}

The initial testing procedure, presented in this section, aims at comparing the specific coefficient estimated with a given value or to compare the values of different coefficients. Both these tests rely on Theorem \ref{Theorem:ConsistanceMLE} to formulate an estimator for the variance of the MLE for the Linear Multidimensional EHP. Although some versions of this test already exist in \cite{dachian2006hypotheses}, they are more limited in scope. Moreover, its theoretical basis is well-established and documented in the scientific literature since \cite{clinet2017statistical}. However, it does not appear in the explicit form of a structured test as proposed here. 

\vspace{\baselineskip}

\begin{test}\label{Test:OneCoefficient}
$\Hz$: $\theta^\star_i=\theta^\star_0$ vs $\Hu$: $\theta^\star_i\neq\theta^\star_0$
 \begin{enumerate}
     \item Compute the MLE $\hat{\theta}$ of $\theta^\star$;
     \item Compute $\hat{I}$ given in Equation \eqref{eq:Ihat};
     \item Compute $Z_i=\frac{\sqrt{T}( \hat{\theta_i} - \theta^\star_{0})}{\hat{\sigma_i} }$, where $\hat{\sigma_i}=\sqrt{(\hat{I}^{-1})_{ii}}$;
     \item Reject $\Hz$ if $ \vert Z_i \vert > q_{1-\alpha/2}$ where $q_{1-\alpha/2}$ is the $1-\alpha/2$ quantile of the standard Gaussian distribution.
 \end{enumerate}
 \end{test}

\vspace{\baselineskip}

This test presents two particularly interesting applications, living outside the theoretical testing framework described earlier. The first one involves assessing the presence of interaction between two sub-processes, indicating a test for $a_{ij}=0$. Another purpose is to evaluate the significance of the mark's impact on the process, meaning testing $\gammaij=0$ for all $i,j \setindex$. However, these specific tests lack theoretical guarantees due to the non-invertibility of the Fisher information matrix in the first case, and the lack of demonstration of the MLE convergence for the second one. Typically, the only case where convergence has been established when $a=0$ is the unidimensional Hawkes process, where the values of $m$ and $b$ are assumed to be known, in \cite{dachian2006hypotheses}.

Nevertheless, this test facilitates the comparison of estimated values of $a$, $b$, and $m$ against predefined values, which can be useful if the literature leads the user to consider a specific value for a parameter of the Hawkes Process. Furthermore, let us highlight that it is possible to generalize this procedure to test the hypothesis $\Hz$: $\theta^\star_i \geq \theta^\star_0$, as the associated test procedure is similar to the one presented above, up to changing the reference quantile from $q_{1-\alpha/2}$ to $q_{1-\alpha}$ and $\vert Z_i \vert$ to $Z_i$.


It is also possible to test equality between two coefficients that determine the Hawkes process. Although the test allows for testing equality between two coefficients associated with a component of the process, and can therefore be used when there is only one sub-process (i.e., for d=1), the major interest of the test lies in testing equality of coefficients between two sub-processes in the multidimensional case $(d>1)$. This test, described just below, relies on the exact same theoretical framework but takes a different form. Consequently, to ensure theoretical guarantees for this test, one must again employ the framework of a Linear Multidimensional EHP.

\vspace{\baselineskip}
\begin{test}\label{test:testdifftheta}
$\Hz$: $\theta^\star_{i}=\theta^\star_{j}$ vs $\Hu$: $\theta^\star_{i}\neq \theta^\star_{j}$
 \begin{enumerate}
     \item Compute the MLE $\hat{\theta}$ of $\theta^\star$;
     \item Compute $\hat{I}$;
     \item Compute $Z_{ij}=\frac{\sqrt{T}( \hat{\theta_i} - \hat{\theta_j})}{ \sqrt{(\hat{I}^{-1})_{ii} - 2(\hat{I_i}^{-1})_{ij} + (\hat{I}_i^{-1})_{jj}} }$;
     \item Reject $\Hz$ if $ \vert Z_{ij} \vert > q_{1-\alpha/2}$ where $q_{1-\alpha/2}$ is the $1-\alpha/2$ quantile of the standard Gaussian distribution.
 \end{enumerate}
\end{test}
%
\vspace{\baselineskip}

This testing procedure serves a crucial role in examining some assumptions commonly applied in Hawkes processes, which posits certain types for the matrix $(\bij)_{i,j \setindex}$. Common assumptions include that the times between events in two sub-processes depend only on the receiving process (meaning in our model that $\bij $ depends only on $i$ and not on $j$) as in \cite{bonnet2023inference}: this assumption could be checked by testing $H_0$: $\bij = b_{ik} $ for all $i$, $j$ and $k$. Another framework, chosen for instance by \cite{deutsch2022bayesian}, assumes each sub-process has the same regulation time and the interaction regulation parameters are the same for all pairs of sub-processes (meaning $\bij = b_1 $ for all $i\ne j$ and $b_{ii} = b_2 $ for all i). This can also be verified by testing first $H_0$: $\bij = b_{kl}$ for all $i\neq j$ and $k \neq l$, and then $H_0: b_{ii} = b_{jj}$ for all $i$ and $j$.

In the context of a Linear Multidimensionnal EHP, this testing procedure offers additional theoretical robustness to this assumption. For instance, if one aims to test a hypothesis of the form $\forall j \neq j', b_{ij}  = b_{ij'}$, this procedure can be adapted by conducting individual sub-tests for each pair of coefficients and rejecting the null hypothesis as soon as any of the subtests rejects it. However, this approach introduces a risk of false negatives, particularly when dealing with numerous sub-processes. To mitigate this risk, it is essential to couple this procedure with a multiple testing correction method, such as \cite{benjamini1995controlling}, \cite{benjamini2001control}, or others to control the overall false positive rate while accommodating the increased risk of false negatives associated with multiple hypothesis testing.
For example, a simpler approach, known as the Bonferroni correction would be to perform the procedure with the desire test level divided by the number of tests performed, meaning replace the level $\alpha$ by $\alpha$ divided by the number of tests to be performed in the procedure Test \ref{test:testdifftheta}.

\subsubsection{Test the mark impact in the infinite time framework: the Z-score}

We present here another test, which also operates within an infinite-time framework, to select the most relevant model between a marked or unmarked model. This test, theoretically described in \cite{clinet2017statistical} or \cite{richards2022score}, differs from Test \ref{Test:OneCoefficient} as it offers to evaluate the coefficient considering that the mark does not impact the process, i.e within the framework of an unmarked model, thereby reducing computation times associated with likelihood maximization. Nevertheless, this test requires the stationarity Assumption \ref{Assumption:stationnarity} to be verified, otherwise risking convergence loss (as, if the assumption is violated, there is no way to see if the mark impacts the process). As previously stated, this condition greatly complicates the form of functions $\phi_\gamma$, as this assumption adds another parameter and the normalization constant can be complex to compute explicitly, depending on the chosen density type. Plus, it still requires an approximation of the Fisher information in the marked case, which can be complex and time-consuming to compute.

\vspace{\baselineskip}

\begin{test}\label{Test:GammaNull}
    $\Hz$: $\forall i,j \setindex ~ \gammaij  = 0$, \textit{v.s.} $\Hu$: $\exists ~(i,j)$ s.t. $\gammaij  \neq 0$.
\begin{enumerate}
\item Compute the MLE $\hat{\theta}^{NM}$ associated with model \eqref{eq:LMEHP};
\item Compute $\w{I}\left(\hat{\theta}^{M}\right)$, a consistent approximation of the Fisher information evaluated at the coefficient $\hat{\theta}^{M}:= (\hat{\theta}^{NM},0_{\mathbb{R}^{d^2}}) $;
\item Compute the vector $\partial_{\gamma} \ell (\hat{\theta}^{M}):= \left(\partial_{\gammaij } \ell (\hat{\theta}^{M})\right)_{1 \leq i,j \leq d} $, where $\ell$ is given by \eqref{theo:LikelihoodMark};
\item Compute the statistic $Z:= \partial{\gamma} \ell (\hat{\theta}^{M})  \hat{I}\left(\hat{\theta}^{M}\right) \partial_{\gamma} \ell (\hat{\theta}^{M})^T $;
\item Reject $\Hz$ if $ Z > q_{1-\alpha}$ where $q_{1-\alpha}$ is the $1-\alpha$ quantile of the chi-square distribution with $d^2$ dimensions.
\end{enumerate}
\end{test}

\vspace{\baselineskip}

The test relies largely on the fact that in absence of marks, i.e., when the $\gammaij $'s are null, the likelihood as well as the Fisher matrix have a simplified expression. As noted in point $2$ of Test \ref{Test:GammaNull}, any consistent approximation of the Fisher matrix induces a functional test with usual theoretical guarantees in infinite time. There are several ways to approximate the Fisher matrix, including $\partial^2_\theta l_T(\theta)$, as discussed in Equation \eqref{eq:Ihat} above.

It is important to note that in point 5. of the test, $Z$ is compared to a chi-square distribution with $d^2$ degrees of freedom, assuming there are $d^2$ mark parameters (one for each interaction between process pairs). However, more generally, different dimensions for mark coefficients can be considered. For instance, assuming all subprocesses are all affected in the same way by the mark (meaning $\gammaij  = \gamma$ for all $i,j$) or that only the receiving process dictates how the mark influences it (meaning $\gammaij  = \gamma_i$). In such cases, the fifth step of the test needs to be adapted to compare $Z$ to a chi-square distribution of dimension $p$, where $p$ is the dimension of the parameterized mark vector.


\subsubsection{Test using a bootstrap procedure}
We describe a test based on a bootstrap procedure proposed by \cite{cavaliere2023bootstrap}. Let us consider the set of points $(T_k)$, from which we estimate the parameter $\theta$ of a Hawkes model thanks to a maximum likelihood approach. This estimated parameter, denoted $\w{\theta}$ is then used to generate $B$ replicates of an inhomogeneous Poisson process with intensity $\lambda_{\w{\theta}}(t)$, the event times of which are denoted $(T_i^{\star, b})$, for each $b=1, \ldots, B$. This intensity is treated as a deterministic function of time, conditionally on the initial data.
From these simulated event times, we re-estimate parameters $\theta^\star_T$ by maximizing the following likelihood $$\ell_T(\theta) =  \sum_{T_i^{\star, b}< T} \log \left( \lambda_\theta( T_i^{\star, b})\right) - \int_{(0,T)} \lambda_\theta(s)ds.$$
Unlike for usual likelihood of a point process, this likelihood both depends on $(T_k)$, the original data, and $(T_i^{\star, b})$. The challenge in this step arises from the evaluation of the intensity $\lambda_{\theta}$, the intensity of the original Hawkes process, at the new jump times and computing the compensator, which depends on the initial data. 
By iterating this procedure  for $1 \leq b \leq B$ with $B$ large enough, we can estimate the variability of the parameters from the empirical variance of $(\theta^{\star,b}_T)_{1\leq b \leq B}$. This variability is expected to be asymptotically equivalent the variability of $\w{\theta}$, thus enabling the construction of a statistical test.

\begin{test}\label{Test:Boots}
$\Hz$: $\theta^\star_i=\theta^\star_0$ vs $\Hu$: $\theta^\star_i\neq\theta^\star_0$
 \begin{enumerate}
     \item Compute the MLE $\hat{\theta}$ of $\theta^\star$;
     \item For $b$ in $\{1, \ldots, B\}$:
     \begin{enumerate}
         \item Simulate an inhomogenous Poisson process with intensity $t \mapsto \lambda_{\w{\theta}_T}(t)$ on $[0,T]$.
         \item Compute the MLE estimator  $\theta^{\star,b}_T$ for the likelihood $\ell_T(\theta) =  \sum_{T_i^{\star, b}< T} \log \left( \lambda_\theta( T_i^{\star, b})\right) - \int_{(0,T)} \lambda_\theta(s)ds.$
     \end{enumerate}
     
     \item Compute $Z_i=\frac{( \hat{\theta_i} - \theta^\star_{0})}{\hat{\sigma_i}^B }$, where $\hat{\sigma_i}^B$ is the empirical standard deviation of the sample $(\theta^{\star,b}_T)_{1 \leq i \leq B}$.
     \item Reject $\Hz$ if $ \vert Z_i \vert > q_{1-\alpha/2}$ where $q_{1-\alpha/2}$ is the $1-\alpha/2$ quantile of the standard Gaussian distribution.
 \end{enumerate}
 \end{test}

This test follows the same theoretical framework as the previous ones, since it is based on the asymptotic normality of the MLE estimator and its consistency property. As a result, the test also requires the MEHP framework to ensure theoretical guarantees. In order to check how this test behave when it falls outside the linear model framework, we implemented it when the model is marked and exhibits inhibition.


\subsection{Test relying on repeated observations of the process}

In the case where multiple repetitions of the process are available, a broader range of tests  is available, including those aimed at evaluating the chosen model type. Indeed, the trajectories of processes can vary significantly from one realization to another, particularly due to the temporal dependency that can affect the process when $\aij$ are non-zero. In such circumstances, assessing the quality of the estimation becomes particularly challenging, especially since, as we will discuss in a later section, using the same sample for both estimation and testing leads to an overestimation of the quality of the estimation.

\subsubsection{Evaluation of the model: the Goodness-of-Fit test}
\label{subsection:GoF}

One of the most widely-used tests associated with the Hawkes process is the Goodness of Fit (GoF) test. In its classical version, which is the most commonly employed when conducting tests on point processes, this test allows for the evaluation of the null hypothesis 
$$\Hz_{,\theta}: "(\Tk)'s \text{ represents a realization of a Hawkes process of intensity } \lambda_\theta". $$
Indeed, thanks to the time-change Theorem \ref{Theorem:TimeChangeMark}, we know that, under the aforementioned null hypothesis, $(\Lambdatheta{}(\Tk))_k$ is a Poisson process with intensity one. Consequently, the sequence $(\Lambdatheta{}(T_{k+1}) - \Lambdatheta{}(\Tk))_k$ follows an exponential distribution with parameter 1. By employing a Kolmogorov-Smirnov (KS) test (or any other test allowing for the comparison of the obtained empirical distribution with a theoretical distribution), we can derive a p-value associated with the null hypothesis.

Let us recall what happens when only one sample of the process is available. In this case, the MLE $\w{\theta}$ can be estimated on the observed process realization, and then injected into the null hypothesis presented above, thus testing the null hypothesis $\ensuremath{\mathrm{H_{0, \w{\theta}_T}}}$. However, this practice has been shown to result in an excessive acceptance of the null hypothesis (see Figure \ref{fig:GoF_resample_procedure} in the appendix). Theoretical explanations for this phenomenon have been explored in \cite{baars2025asymptotically}. Thus, when using only one realization of the process, we are unable to assess the quality of the estimation made via maximum likelihood with this test. Following usual learning methods, another possibility would be, when having access to at least two independent repetitions of the process, to perform the estimation and the test on the different samples available. However, this technique has also shown its drawbacks, as it leads to an underestimation of the associated p-values of the KS test and thus leads to rejecting the null hypothesis more frequently than it should (see Figure \ref{fig:GoF_resample_procedure} in appendix).
To address these issues, \cite{reynaud2014goodness} propose, in the case where multiple realizations of the process are available, a procedure of subsampling to obtain, under the null hypothesis, p-values that indeed follow a uniform distribution. This test has the advantage of not specifically testing values of coefficients defining the model, but rather the type of model selected to describe the data. As a result, the theoretical guarantees of this test rely only on the fact that, if the model was correctly selected, the estimation enables a good reconstruction of the given intensity of the process.

We present here an extension of this test that use a resampling procedure to have access to the distributions of the p-values associated rather than a unique p-value. This way, we are allowed to compare several models by looking for the distribution that is more likely to be a uniform distribution

\vspace{\baselineskip}

\begin{test}\label{test:goodness}
$\Hz$: $((T^{(r)}_k)_k)_{1 \leq r \leq n}$ represent $n$ i.i.d. realizations of a point process with intensity $\lambda_{\theta^*}$ on $[0,T_{\max}]$, with $\theta^*$ an unknown parameter.
 \begin{enumerate}
 \item Fix $p(n) \in \{1, ..., n \}$ such that $p(n)/n \xrightarrow[n \to \infty]{} 0 $;
 \item For each $r \in \{1,...,n\}$, compute the MLE $\hat{\theta}^{(r)}$ associated with the realization $(T_{k}^{(r)})$ and compute $\hat{\theta}$ the mean value of $(\hat{\theta}^{(r)})_r$;
 \item For any subset $S$ of $\{1,...,n\}$ of cardinality $p(n)$: 
 \begin{enumerate}

     \item For all $r \in S$, compute $\hat{\mathcal{N}}^r:= \{ \hat{\Lambda}_i(T^{(r)}_k), 1 \leq k \leq \mathcal{N}^r_{T_{\max}} \}$ where $\hat{\Lambda}_r(s) = \int_{ (0,t)} \left(\lambda_{\hat{\theta}}\right)_r(s)ds $ and 
     $(\lambda_{\hat{\theta}})_r$ is the intensity of a Hawkes process of parameter $\hat{\theta}$ and having the arrivals times $(T^{(r)}_k)$;
     \item Compute $\hat{\mathcal{N}}^{\text{cum}}$ the cumulated process associated with $(\hat{\mathcal{N}}^r)_r$ by recurrence on elements of $S:= \{ r_1, ..., r_{p(n)} \}$:
     
     \begin{itemize}
         \item \text{ Initialization, $j=1$: }$\hat{\mathcal{N}}^{\text{cum}} \xleftarrow[]{} \hat{\mathcal{N}}^{r_1}$;
         \item \text{ Recurrence on $1 \leq j \leq p(n) $: }$\hat{\mathcal{N}}^{\text{cum}} \xleftarrow[]{} \hat{\mathcal{N}}^{\text{cum}} \cup \left(   \sum_{l<j} \hat{\Lambda}_{r_l}(T_{\max}) + \hat{\mathcal{N}}^{r_j} \right)$; 
     \end{itemize}
     
     \item Choose $\xi >0 $ such that $\xi p(n) < \sum_{r \in S} \hat{\Lambda}_r(T_{\max})$ almost surely;
     \item  Perform a test to compare the distribution of the cumulative process $ \left( \frac{1}{\xi p(n)}\hat{\mathcal{N}}^{\text{cum}}_{t} \right)_{t \in [0, \xi p(n)]}$ to a uniform distribution on $[0,1]$ and recover the associated p-value $p_S$; 
 \end{enumerate}
 \item  Compare $(p_S)_S$ to a uniform distribution
 \end{enumerate}
\end{test}

\vspace{\baselineskip}

While the true parameter value $\theta^*$ is not directly involved in the test, it plays an implicit role. For the expression $\frac{1}{\xi p(n)}\hat{\mathcal{N}}^{\text{cum}}$ to converge to a uniform distribution, the parameter estimator $\hat{\theta}$ must satisfy $p(n)^{-1/2} \left[ \sum_{i \in S } \Vert \left(\lambda_{\hat{\theta}}\right)_i - (\lambda_{\theta^*})_i \Vert_{l_1([0;T_{\text{max}}])} \right] \xrightarrow[n \to \infty]{\mathbb{P}} 0$. Therefore, the only assumptions for this test pertain to the chosen model type and the quality of the estimation made under the null hypothesis, as indicated by the probability convergence mentioned earlier. However, this condition as not been proven for a Linear EHP whether marked or not, as the convergence rate has not been studied in the asymptotic framework where we consider an increasing number of repetitions on the same interval of observations.

The model type under examination is reflected in the form of the compensator, which is directly derived from the chosen intensity function. Each model induces a distinct compensator, resulting in different p-values. For example, let us consider two different models yielding distinct compensators $\Lambda_1$ and $\Lambda_2$. By employing the described test, one can obtain two sets of p-values $(p_{S,1})$ and $(p_{S,2})$. These distributions can then be used to determine the most appropriate model by evaluating which one closely resembles a uniform distribution. However, there is no universal criterion for such a test. Various criteria can be considered; for instance, one may count the number of rejections associated with each $(p_{S,i})$ and select the model with the least rejections. Alternatively, one could compute the dissimilarity between each $(p_{S,i})$ and a uniform distribution, then select the one that is the most similar to the uniform distribution. This procedure can be extended to compare more than two types of models. It is worth noticing that any valid test can be used to assess the point $3$ of Test \ref{test:goodness}. The most famous test in this type of situation is usually the Kolmogorov-Smirnov test, but other tools are available (see in Section \ref{Section:Simulation} the qqconf plots).

In cases where only one model is of interest, it is possible to streamline the process by omitting the bootstrap procedure and simply use a single sample $S$ randomly selected from all possible subsets of size $p(n)$. In this scenario, only the p-value associated with the test comparing $ (\xi p(n))^{-1}\hat{\mathcal{N}}^{\text{cum}}$ to a uniform distribution needs to be considered to reject $\Hz$. 

It can also be noted that item 4 suggests testing whether the points of the process are identically distributed over a given time interval to verify if the points of the cumulative process come from a Poisson process with unit intensity. Another technique would be to test whether the increments of this same process follow an exponential distribution with a unit parameter. However, simulations using this second method have shown that the exponential test required more points to produce p-values following an exponential distribution. For this reason, we have opted to keep the test on the uniform distribution of points.

\subsubsection{Compensator for different models}

As highlighted earlier, the compensator assumes a pivotal role in evaluating the suitability of the chosen model in light of the data as its precise computation also provides access to the likelihood function and, consequently, to the maximum likelihood estimator. This section is dedicated to exploring the diverse forms that the compensator can adopt across the various models associated with the Hawkes Process. We consider, in this section, models allowing for excitation as well as models allowing for inhibition.

Compensator computation in the non-linear case requires an additional assumption concerning the temporal regulation parameters defining the process. Although this assumption is not mandatory in the linear case, it greatly simplifies the intensity calculations and is therefore also sometimes used in this context. 

\begin{ass}
\label{Assumption:Equality_Betaij}
In Model \ref{def:NLMMEHP} , we assume that, for all $i \setindex $, there exists $b_i >0$ such that for all $j \in { 1, \ldots, d } $, $b_{ij}  = b_i$.
\end{ass}

This assumption impacts how we consider the interactions between our different sub-processes. Supposing that $\bij $ depends only on $i$ implies that the typical duration of the interaction between the two sub-processes $i$ and $j$ depends only on the receiving sub-process and not on the emitter-receiver pair. From a practical standpoint, this ensures that the intensity can be expressed as a decreasing exponential and not as a sum of decreasing exponential with different decay rates, thus easing the computation.

In scenarios where the coefficients $\aij$ are assumed to be strictly positive, computing the compensators is relatively straightforward, as the intensity function never reaches zero. These computations have been well-established for some time now in \cite{laub2021elements}. However, when non-linearity is introduced in the model via $\aij<0$, computing the compensator becomes significantly more complicated as it requires identifying the intervals of time when the intensity is non-zero.

\paragraph{Linear process.} For linear cases, the computation of the compensator has been extensively addressed in the literature, and expressions for it are readily available. Using the results in \cite{laub2021elements}, we present here the detailed expressions. 

\begin{prop}
    \label{Proposition:CompensateurLinear}
    Consider a Linear Multidimensional Marked EHP  (resp unmarked) and denotes $( \Tk, m_k, \kappa_k)$  (resp $( \Tk, m_k)$)  the associated times and component and mark. The compensator associated with the process is given by 
    \begin{equation}
    \label{eq:CompensatorLinear}
        \Lambdatheta{}(T) = \sum_{i=1}^{d} \left(  m_i T_{1} +   \sum_{k=1}^{N(T)}  m_i \left( T_{k+1} - T_{k} \right) + b_i^{-1} \left( \lambdatheta{i}( T_{k}^+) - m_i \right)\left( 1 - e^{- b_i \left( T_{k+1} - T_{k} \right)} \right)\right) 
    \end{equation} where:
    \begin{enumerate}
        \item $\lambdatheta{i}( T_{k}^+) = \lambdatheta{i}( T_{k}) + a_{i m_k} \phi_{\gamma_{im_k}}(\kappa_k)$ in the marked case,
        \item  $\lambdatheta{i}( T_{k}^+) = \lambdatheta{i}( T_{k}) + a_{i m_k}$ in the unmarked case,
    \end{enumerate}
and  $\lambdatheta{i}( t^+) = \lim_{ s \to t~ s > t } \lambdatheta{i}(s)$.
\end{prop}

\paragraph{Non-linear process.}

To compute the compensator in the non-linear case, we first need to identify the intervals of time on which the intensity is zero. Using this same idea, initially introduced in  \cite{bonnet2021maximum} for a multidimensional Hawkes process without marks, we present the expression of the compensator associated in the cases with the mark.

For all $i \in \{1,...,d\}$ and for all $ k \geq 0$, $T_{k}^{i,*}$ denotes the smaller integer between $\Tk$ and $T_{k+1}$ such that the intensity of the $i$-th sub-process is strictly positive, meaning 

$$T_{k}^{i,*} = \inf \left(  T_{k+1}, \inf \{ t \geq  T_{k}  \text{ }: \text{ } \lambdatheta{i,\star}(t) \geq 0\} \right).$$

\begin{prop} 
    \label{Proposition:CompensateurNonLinear}
Considering a non-linear Hawkes process, the expression of the compensator is given by  
\begin{equation}
\label{eq:CompensatorNonLinear}
    \Lambda_{\theta}(T) = 
          \sum_{i=1}^{d}  m_i T_{1} +  \left(\sum_{k=1}^{N(T)} J_k \right)
\end{equation}
with 
\begin{align*}
        &T^{i,*}_{k}  = \min \left(T_{k}  + b_i^{-1} \log \left( \frac{m_i - \lambdatheta{i,\star}(\Tk^+) }{m_i} \right) \indic_{ \lambdatheta{i,\star}(\Tk^+)<0 }, T_{k+1} \right), \\
        &J_k   = m_i \left(  \min(T, T_{k+1})-  T_{k}^{i,*} \right) + b_i^{-1} \left( \lambdatheta{i,\star}(\Tk^+)- m_i \right)\left( e^{ -b_i (T_{k}^{i,*}- T_{k})} - e^{ -b_i \left( \min(T, T_{k+1})-T_{k} \right)} \right), 
    \end{align*} and 
    \begin{enumerate}
        \item $\lambdatheta{i,\star}( T_{k}^+) = \lambdatheta{i,\star}( T_{k}) + a_{i m_k} \phi_{\gamma_{im_k}}(\kappa_k)$ in the marked case,
        \item  $\lambdatheta{i,\star}( T_{k}^+) = \lambdatheta{i,\star}( T_{k}) + a_{i m_k}$ in the unmarked case,
    \end{enumerate}
where $\lambdatheta{i,\star}( t^+) = \lim_{ s \to t~ s > t } \lambdatheta{i,\star}(s)$.
    
\end{prop}
The demonstration of this proposition in the unmarked case can be found in \cite{bonnet2023inference}. We provide a demonstration for the marked case in the appendix.

As highlighted by Propositions \ref{Proposition:CompensateurLinear} and \ref{Proposition:CompensateurNonLinear}, the expressions in the marked or unmarked cases are almost identical. Upon comparison, we observe that the only modification when adding the mark lies in the value of $\lambdatheta{i}(\Tk^+)$, with modifications from $\lambdatheta{i}(\Tk^+)+ a_{im_k}$ to $\lambdatheta{i}(\Tk^+) + a_{im_k}\phi_{\gamma_{im_k}}(\kappa_k)$ in the marked scenario. This is linked to the fact that the mark only impacts the interaction between subprocesses at the arrival times, and as a result, only changes the height of the jump made by the intensity. Consequently, the behavior of the underlying intensity is very similar with or without marks.

\section{Experiments: synthetic data}
\label{Section:Simulation}

The aim of this section is to evaluate the different test procedures presented in Section \ref{Section:Test_Procedure}, particularly when moving away from the theoretical reference framework. Since this numerical study is already extensive, we only present the results in such settings and not those obtained within the classic theoretical framework. However, in order to provide a clear overview of the simulations carried out, the following Table \ref{tab:bigtable} summarizes all the simulations and tests performed.

\begin{table}
    
{ \footnotesize \centering
\begin{tabular}{ |C{0.15\linewidth}|C{0.2\linewidth}|C{0.07\linewidth}|C{0.35\linewidth}|C{0.1\linewidth}| }
\hline
 Model &  Model for estimation & Test applied & Hypothesis $\Hz$ under test & Theoretical guarantees \\
\hhline{|=|=|=|=|=|}
   \multirow{3}*{ \hfill Poisson model } &  Poisson model &Test \ref{Test:OneCoefficient} & $a^*=0$ &  No 
\\\cline{2-5}
  & Hawkes Model &Test \ref{Test:Boots} & $a^\star = 0$ & No
   \\\cline{2-5}
  & Poisson Model &Test \ref{test:goodness} & Poisson Process & Yes  \\\cline{2-5}
 & Linear EHP &Test \ref{test:goodness} &  Hawkes Process  & No   
  \\\hhline{|=|=|=|=|=|}
 \multirow{5}{*}{ \parbox{\linewidth}{Linear EHP  with d=1} } & Linear EHP & \multirow{2}{*}{ Test  \ref{Test:OneCoefficient}  } & $a^* = a_0 >0$  &  \multirow{3}{*}{Yes}   \\
 &&and & $b^* = b_0 >0$  & \\
 && Test \ref{Test:Boots} & $m^\star = m_0 >0 $& \\\cline{2-5}
 &Linear Marked EHP& Test \ref{Test:OneCoefficient} & $\gamma^* =0 $ & No  \\\hhline{|=|=|=|=|=|}
  
\multirow{4}{*}{ \hfil \parbox{\linewidth}{Linear Hawkes with d>1} } &Linear EHP   & \multirow{3}{*}{ Test  \ref{Test:OneCoefficient}} & $a_{ij}^* = a_{ij} >0$  &  \multirow{3}{*}{Yes}   \\
 && & $b_i^* = b_i >0$  & \\
 &&  & $m_i = m_{i} >0 $& \\\cline{2-5}
& Linear EHP & Test \ref{test:testdifftheta} & $\theta_i^* = \theta_j^*$ &  Yes \\\hhline{|=|=|=|=|=|}

 \multirow{4}{*}{ \centering \parbox{\linewidth}{Non-Linear EHP  with d=1} } & Non-Linear EHP & \multirow{3}{*}{ Test  \ref{Test:OneCoefficient}} & $a = a^* <0$  &  \multirow{3}{*}{No} \\
 && & $b = b^* >0$  & \\
 &&  & $m = m^* >0 $&  \\\cline{2-5}
 &Non-Linear EHP& Test \ref{test:goodness} &  Hawkes Process with $a^*<0 $ & No  \\\cline{2-5}
  &Linear EHP& Test \ref{test:goodness} &  Hawkes Process with $a^*>0 $ & No  \\\hhline{|=|=|=|=|=|}
  
 \multirow{5}{*}{ \parbox{\linewidth}{Linear Marked EHP  with d=1} } & Linear Marked  EHP & \multirow{2}{*}{ Test  \ref{Test:OneCoefficient}} & $a = a^* >0$  &  \multirow{4}{*}{No}  \\
 &&and& $b = b^* >0$  & \\
 && Test \ref{Test:Boots}& $m = m^* >0 $&  \\
 &&& $\gamma = \gamma^* \ne 0$  & \\\cline{2-5}
 &Linear EHP& Test \ref{test:goodness} &  Hawkes Process  & No  \\\cline{2-5}
  &Linear EHP& Test \ref{test:goodness} &  Marked Hawkes Process & No  \\\cline{2-5}
  & Linear EHP & Test \ref{Test:GammaNull} & $\gamma^*=0$ & No \\
  \hhline{|=|=|=|=|=|}

 \end{tabular}
 }
\caption{Synthetic presentation of the testing procedures.} 
\label{tab:bigtable}
\end{table}

Let us recall that the GoF procedure is not supported by theoretical guarantees as long as the property $p(n)^{-1/2} \left[ \sum_{i \in S } \Vert \left(\lambda_{\hat{\theta}}\right)_i - (\lambda_{\theta^*})_i ||_{l_1([0;T_{\text{max}}])} \right] \xrightarrow[n \to \infty]{\mathbb{P}} 0$ has not been proven, which is, to the best of our knowledge, the case for any type Hawkes Process under consideration. Indeed, this hypothesis assumes a particular behavior for the MLE when the number of repetitions increases, whereas all theoretical results on Hawkes use a theoretical framework in which it is the observation time of the process that tends towards infinity, and not the number of repetitions. The only exception is the Poisson process, for which the MLE estimator can be expressed as the number of points divided by the total observation time, which greatly simplifies the calculations and makes it possible to prove this result.

In addition, Test $\ref{Test:OneCoefficient}$ offers theoretical guarantees only when the simulated processes and the estimation are done considering a Linear Hawkes model with parameter $a^\star>0$. As a result, most cases outlined in the subsequent section exceed the assured theoretical framework, thereby assessing the practical usability of the procedure despite the current lack of theoretical guarantees.

\subsection{Implementation}

The simulation and estimation of a multidimensional marked Hawkes process required the development of several algorithms, available through a git repository: \url{https://github.com/Msadeler/marked_exp_hawkes}.
Moreover, to ensure that the present study is reproducible, a notebook named \texttt{example} is also available containing almost all the tests presented below.
Extending the initial code available in \cite{migmtzGit}, these functions allow for the simulation and estimation of both marked and unmarked Exponential Hawkes Processes (EHP) with either excitation or inhibition mechanisms. The following list provides a summary of the key functions and their purposes.

\begin{itemize}
    \item \texttt{exp\_thinning\_hawkes\_marked} and  \texttt{multivariate\_exponential\_hawkes\_marked}: simulates an unmarked or marked EHP using the thinning algorithm, incorporating both excitation and inhibition.
    \item \texttt{estimator\_unidim\_multi\_rep} and \texttt{estimator\_multidim\_multi\_rep}: estimates parameters of a unidimensional or multidimensionnal EHP when multiple repetitions of the process are available.
    \item \texttt{loglikelihood\_estimator\_bfgs} and \texttt{multivariate\_estimator}: estimates parameters of an MEHP when only one repetition is available.
\end{itemize}

\subsection{Simulation scheme and evaluation}

This section aims at providing the general framework within which our simulations and estimations are conducted. We first present the scenarios chosen to generate the data, and then explain how we evaluate the different test procedures.

\subsubsection{Simulation scenario}
\label{subsubsection:DataGeneration}


For each test, we generate $n = 500$ independent and identically distributed repetitions of the process, each within the interval $[0, T_{\max}]$ with $T_{\max} = 5000$. Each type of process is generated with constant parameters throughout the simulations:
\begin{itemize}
    \item  Poisson processes are generated with an intensity of $m^\star = 1$;
    \item  Linear  Unidimensional Hawkes processes are generated with $m^\star = 1$, $a^\star = 0.6$, and $b^\star = 2$;
    \item Non-Linear Unidimensional ($d=1$) Hawkes processes are generated with $m^\star = 1$, $a^\star = -0.2$, and $b^\star = 2$. For this particular model, simulations are done using $T_{\max} = 20000$;
    \item Linear Unidimensional ($d=1$) Marked Hawkes processes are generated taking $m^\star = 1$, $a^\star = 0.6$ $b^\star = 2$ and $(\kappa_i)$ following, an exponential distribution with parameters $\psi^\star = 0.5$, with either $\phi_{\gamma}(x) = e^{\gamma x}$ or $\phi_{\gamma,\psi}(x) = \frac{\psi - \gamma}{\gamma} e^{\gamma x}$ depending on whether the simulations are conducted considering Assumption \ref{Assumption:stationnarity}.
    \item  Multidimensional Hawkes processes are generated with two dimensions, i.e $d=2$, with $m^* = \begin{pmatrix} 0.5 & 0.2 \end{pmatrix}^T$, $a^* = \begin{pmatrix} 0.4 & 0.2 \\ 0.2 & 0.6 \end{pmatrix}$, and either $b^* = \begin{pmatrix} 1 & 1 \end{pmatrix}^T$ or $b^* = \begin{pmatrix} 1 & 1.5 \end{pmatrix}^T$ depending on the null hypothesis tested. 
\end{itemize}

\subsubsection{Evaluation of the test procedure}
\label{subsubsection:EvaluationTest}

In this section, we present the various procedures used to assess test performance. These procedures, which vary according to the type of test under consideration, enable us to assess the Type I and Type II risks of different tests.

\paragraph{Goodness-of-fit tests.} To assess the practical performance of Test \ref{test:goodness}, the main idea is to investigate the ability of the test to detect whether the model is accurately specified or not. Therefore, there is one model under which the data is generated, and one model used for estimation, both of which can be the same (in that case, the model is well specified) or different (in that case, the model is mispecified). More precisely, we generate data under a model $M_{data}$ that can be either a Poisson process, a linear Hawkes process, a nonlinear Hawkes process or a linear marked Hawkes process.
Then, we compute the MLE under another model $M_{estimator}$, that belongs to the same class of models than $M_{data}$ but is not necessarily the same one. Then, our goal is to determine whether the model $M_{estimator}$ fits accurately the data according to the goodness-of-fit test described in Test \ref{test:goodness}.
Since the latter contains a resampling step, the output of the procedure is a sequence of p-values, that shall be compared to a uniform distribution. This final step is achieved thanks to the \texttt{qqconf} R package that is described at the end of this subsection.

\paragraph{Bootstrap tests.} The bootstrap procedure does not provide direct estimators with an expected distribution that could be verified. However, it does provide empirical confidence intervals that can be used to assess the validity of the procedure. Therefore, we focus on the test rejection rate (i.e. the proportion of times when the true parameter falls outside the estimated confidence intervals) when simulations are run under the null hypothesis, to check whether it matches the $\alpha$ significance level or not. Let us highlight that this approach is an indirect way of validating the procedure as, although it offers an indication of the test's performance, it does not guarantee a comprehensive assessment of its reliability.

\paragraph{Tests on coefficients.} The evaluation of Test \ref{Test:OneCoefficient} follows the same idea: we simulate i.i.d. samples under $\Hz$ and then under $\Hu$ and we assess both ability of the test to conserve and reject $\Hz$ when appropriate.

Since the different procedures are derived from asymptotic normality results, the validity of which is not always ensured, we also compare the empirical distributions of the assumed asymptotically normal quantities with the Gaussian distribution. 
In order to consider normalized quantities, we employed the empirical standard deviation of the sample instead of the estimator of the Fisher information to avoid issues related to numerical computation and to prevent cases where the Fisher matrix was singular. Nevertheless, the test in the simulation presents the same theoretical guarantees as Test \ref{Test:OneCoefficient}. 


\paragraph{Testing the uniformity or normality assumption.}
For all proposed test procedures, the final step requires either to test a uniformity assumption or a normality assumption. Although this can be done thanks to standard tests such as Kolmogorov-Smirnov or Shapiro-Wilk,  we use here the alternative approach developed and implemented in the R library \texttt{qqconf} \citep{weine2023application}. This library offers a framework for constructing confidence intervals tailored to the empirical distribution, ensuring an evaluation encompassing both the central region and the tails of the distributions. Those confidence intervals are represented in a QQ-plot with an area surrounding the theoretical distribution tested. Notably, this approach offers a compelling alternative by providing a unified framework for all distribution types implemented in R, diverging from conventional testing methods that often concentrate on either the central tendencies or the extreme values of the distributions. 


\subsection{Results on synthetic data}

In this section, we display and comment the qqconf plot derived from the simulation of the processes described in Section \ref{subsubsection:DataGeneration} and estimated using the procedure described in Section \ref{subsubsection:EvaluationTest}.

\subsubsection{Testing the self-exciting assumption.}
In this section, we aim at testing the self-exciting assumption, that is distinguishing between a Hawkes process and a Poisson process. For this purpose, we compare the performances of a test on coefficients (Test \ref{Test:OneCoefficient}) the goodness-of-fit test (Test \ref{test:goodness})  and the bootstrap test (Test \ref{Test:Boots}).

We first investigate the performance of Test \ref{Test:OneCoefficient} with null assumption $\Hz: a^\star=0$. Let us highlight that theoretical guarantees are not ensured here, since the asymptotic normality of the MLE estimators is proven only when $a^\star >0$. Nevertheless, we can verify if the empirical distribution of $\hat{a}$ is close to a Gaussian distribution, which would be a first empirical check.

\begin{figure}[!hbtp]
    \centering
    \includegraphics[scale=0.45]{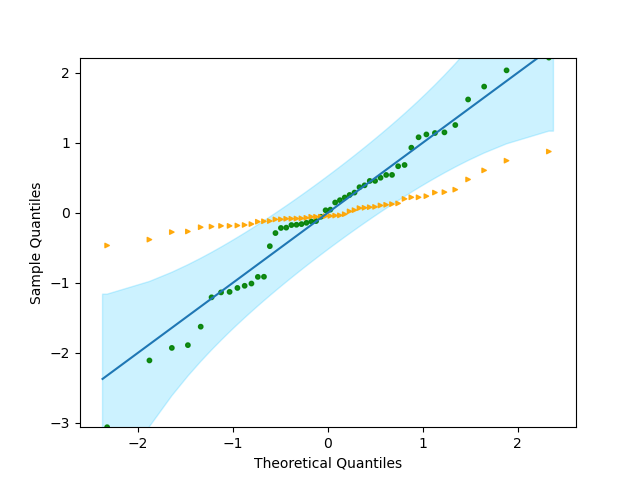}
        \caption{Test \ref{Test:OneCoefficient} - Theoretical quantiles for the standard Gaussian distribution (blue line) with 95\% confidence intervals (blue region) compared to the empirical quantiles of $\frac{\w{a}}{\w{\sigma}}$ when either $\mu^\star$ and $b^\star$ are fixed and known parameters (green points) or when they are estimated (orange points). The data are generated under $H_0$: $a^\star=0$, with $m^\star = 1$, $T_{\max} = 5000$.} 
        \label{fig:Test_Alpha_Null_Hawkes}
\end{figure}

Figure \ref{fig:Test_Alpha_Null_Hawkes} shows that under the null hypothesis $\Hz$, the empirical distribution of $\hat{a}$ is significantly different from a Gaussian distribution, which suggests that Test \ref{Test:OneCoefficient} should not be used in practice, and that no theoretical guarantees can be developed for it. However, the empirical distribution of $\w{a}$, displayed in Figure \ref{fig:Test_Alpha_Null_Hawkes}, is close to a Gaussian when both parameters $m^\star$ and $b^\star$ are known, which is consistent with the theoretical grounds obtained by \cite{dachian2006hypotheses} in this particular case.

\begin{figure}[hbtp]
\centering
        \includegraphics[scale=0.45]{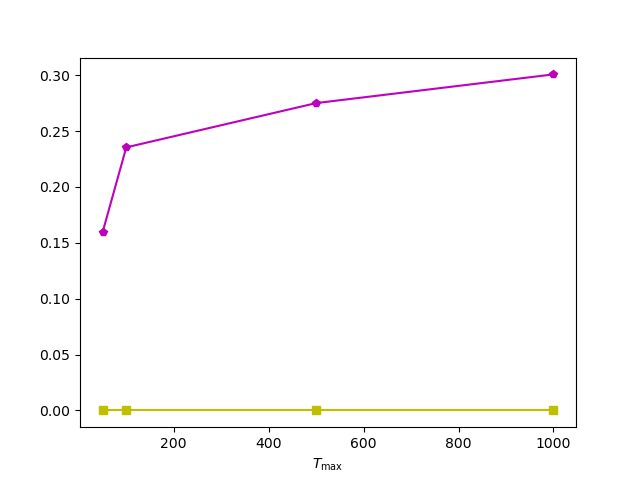}
        \caption{Test \ref{Test:Boots} - Type I (yellow curve) and type II risks (magenta curve) for the bootstrap test. The type II risk is evaluated through simulations of the Hawkes process with $m^\star = 1$ $a^\star=0.6$, $b^\star =1$ and $T_{\max}=5000$.}
        \label{fig:Test_Alpha_Null_Boot}
\end{figure}

Regarding the bootstrap procedure, the type I risk curve as a function of the observation time $T_{\max}$, displayed in Figure \ref{fig:Test_Alpha_Null_Boot}, shows that it fails to achieve a rejection rate of $5 \%$ when supposed to. However, when $a^\star$ is non-zero, the test is able to distinguish between a Hawkes process and a Poisson process.

We then explore whether the goodness-of-fit procedure can allow choosing between the Poisson and the Hawkes assumptions. To that end, we compute the compensators in the Poisson model and in the Hawkes model, and we perform the goodness-of-fit Test \ref{test:goodness} in each scenario. Let us recall that the compensator for the Poisson Process is obtained through Equation \eqref{eq:CompensatorLinear}, when $a=0$.

\begin{figure*}[!hbtp]
    \centering
    \begin{subfigure}{0.47\textwidth}
    \includegraphics[width =\textwidth]{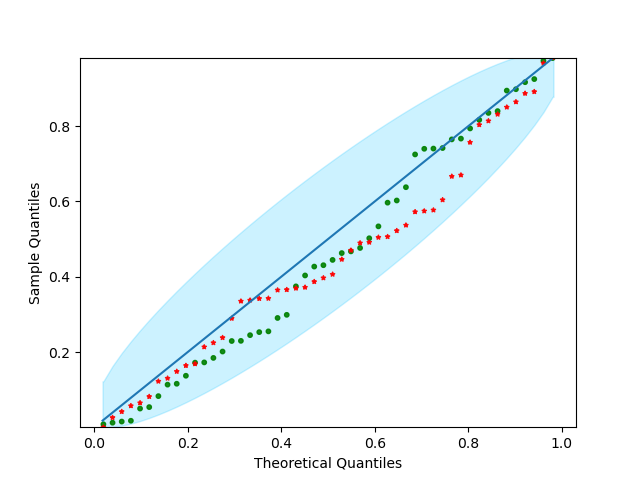}
    \caption{Simulation of a Poisson process}
    \label{fig:GoF_poisson_vs_hawkes_simulation_poisson_risktype_1}
    \end{subfigure}
    \quad 
    \begin{subfigure}{0.47\textwidth}
    \includegraphics[width =\textwidth]{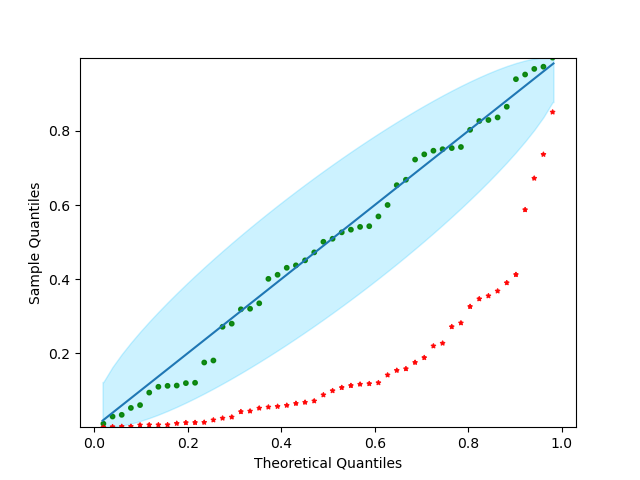}
    \caption{Simulation of a Hawkes process}
    \label{fig:GoF_Hawkes_vs_Poisson_simu_poisson_risktype_2}
    \end{subfigure}
    \caption{Test \ref{test:goodness} - Theoretical quantiles of the Uniform distribution (blue line) with 95\% confidence intervals (blue region), compared to empirical quantiles of the distribution of GoF-based p-values, when simulating either a Hawkes Process with parameters $m^\star = 1$, $a^\star= 1$, $b^\star= 1$ (Figure \ref{fig:GoF_Hawkes_vs_Poisson_simu_poisson_risktype_2}) or a Poisson process with parameters $m^\star =1$ (Figure \ref{fig:GoF_poisson_vs_hawkes_simulation_poisson_risktype_1}). In both cases, estimation and testing are carried out considering either a Hawkes model (green points) or a Poisson model (red points). For all simulation we used $n=500$ repetition and $T_{\max}=5000.$}    
    \label{fig:gof_poisson_vs_hawkes}
\end{figure*}

\begin{figure*}[!hbtp]
\centering
    \includegraphics[scale=0.5]{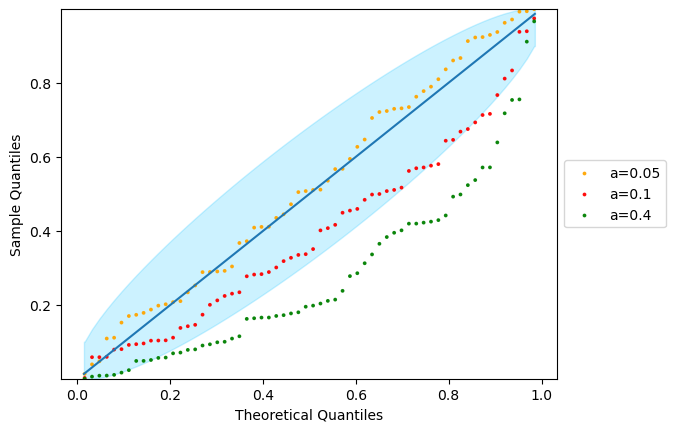}
    \caption{Test \ref{test:goodness} - Theoretical quantiles of the Uniform distribution (blue line) with 95\% confidence intervals (blue region), compared to empirical quantiles of empirical quantiles of the GoF-derived p-values when simulating a Hawkes Process with parameters $m\star = 1$, $a^\star \in \{ 0.05, 0.1, 0.4 \}$, $b^\star= 2$ and $T_{\max}=5000$, $n=500$ and estimating a Poisson process.} 
    \label{fig:GoF_Hawkes_vs_Poisson_simu_poisson_differenta}
\end{figure*}

As shown in Figure \ref{fig:GoF_poisson_vs_hawkes_simulation_poisson_risktype_1}, when a Poisson process is simulated, the two goodness-of-fit tests give similar results. However, when a Hawkes process is generated with $a = 0.6 $, the GoF procedure strongly rejects the Poisson model hypothesis (see Figure \ref{fig:GoF_Hawkes_vs_Poisson_simu_poisson_risktype_2}). Nonetheless, Figure \ref{fig:GoF_Hawkes_vs_Poisson_simu_poisson_differenta} shows that this result strongly depends on the value of the parameter $a $, and even on the ratio $a/b $. Indeed, for $b = 1 $, the figure shows that the Poisson model hypothesis is not rejected when $a $ is low. This suggests that the GoF can allow to distinguish accurately between a Hawkes and a Poisson process, but only if the self-exciting effect is strong enough.

\subsubsection{Test of the inhibition assumption}
\label{subsec:numerical_test_inhibition}
This section aims at assessing the performance of Tests $\ref{test:goodness}$ and $\ref{Test:OneCoefficient}$ to determine if the observed data exhibits inhibition, that can only be accurately detected using a nonlinear Hawkes process. Once again, this scenario falls out of the theoretical guarantees described in Section \ref{Section:TheoreticalProperty}. For this purpose, we display the empirical law associated with each MLE coefficient when the process is simulated with inhibition and compare it to a normal law, as this property, that still remains unproven, is the cornerstone of Test \ref{Test:OneCoefficient}. 

\begin{figure}[!hbtp]
    \centering
    \begin{subfigure}{0.31\textwidth}
    \includegraphics[width =\textwidth]{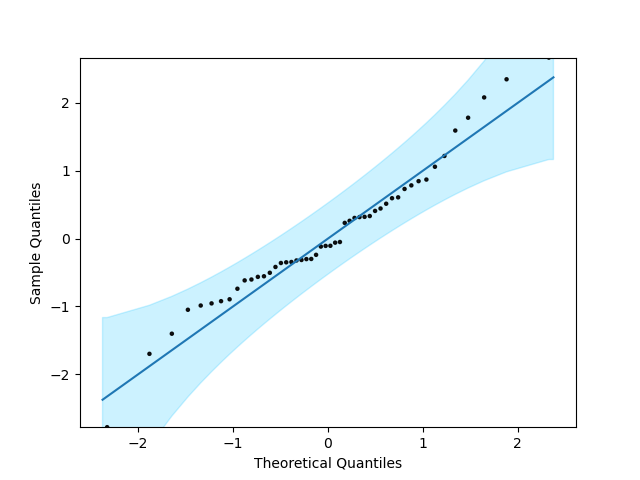}
    \caption{Law of $\w{m}$.}
    \label{fig:Asymptotic_law_mu}
    \end{subfigure}
    \quad 
    \begin{subfigure}{0.31\textwidth}
    \includegraphics[width =\textwidth]{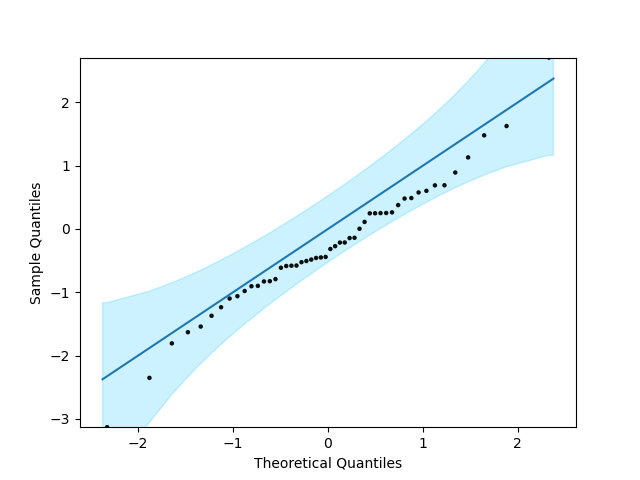}
    \caption{Law of $\w{a}$.}
    \label{fig:Asymptotic_law_alpha}
    \end{subfigure}
    \begin{subfigure}{0.31\textwidth}
    \includegraphics[width =\textwidth]{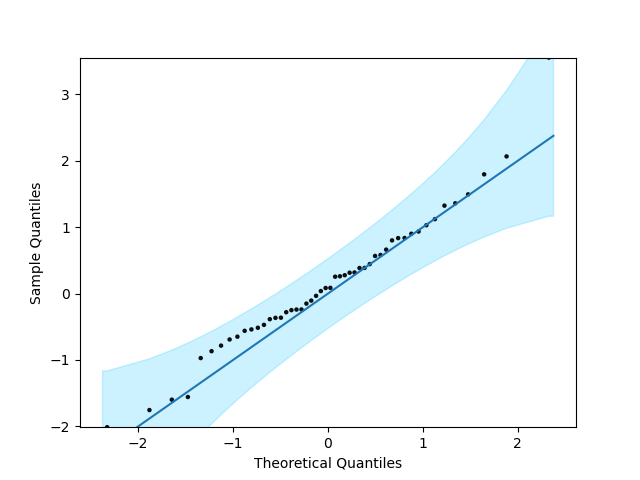}
    \caption{Law of $\w{b}$.}
    \label{fig:Asymptotic_law_beta}
    \end{subfigure}
    \caption{Test \ref{Test:OneCoefficient} - Theoretical quantiles of the standard Normal distribution (blue line) with 95\% confidence intervals (blue region), compared to empirical quantiles of the distribution $\frac{\w{\theta}_i-\theta_i^\star}{\w{\sigma}_i}$ when simulating a Hawkes Process with parameters $m^\star=1$, $b^\star=2$, and $a^\star=-0.6$, $T_{\max}=5000$ and $n=500$ repetitions.}
    \label{fig:Asymptotic_law_with_inhibition}    
\end{figure}

As displayed in Figure \ref{fig:Asymptotic_law_with_inhibition}, asymptotic normality is empirically reached for all parameters. However, achieving this convergence requires a considerable amount of time ($T_{\max}=20000$ for this simulation): as inhibition leads to an intensity function that frequently hits zero, a longer duration is needed to ensure a good estimation and MLE's convergence. Nonetheless, this asymptotic normality suggests that Theorem \ref{Theorem:ConsistanceMLE} could be extended to Non-Linear Hawkes Process.

We then compare this test to the GoF bootstrap procedure. As the compensators associated with a self-exciting or self-inhibiting model are the same as long as $\w{a}$ is positive, we present only the output of the procedure when a self-inhibiting process is simulated.  

\begin{figure}[!hbtp]
    \centering
    \begin{subfigure}{0.41\textwidth}
    \includegraphics[width =\textwidth]{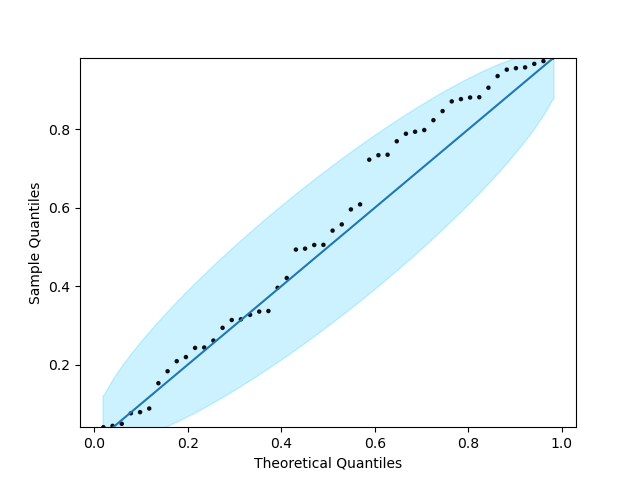}
    \caption{Compensator $\Lambda$ computed using the Nonlinear model \eqref{eq:CompensatorNonLinear}}
    \label{fig:QQconf_inhib_risktype1}
    \end{subfigure}
    \quad 
    \begin{subfigure}{0.41\textwidth}
    \includegraphics[width =\textwidth]{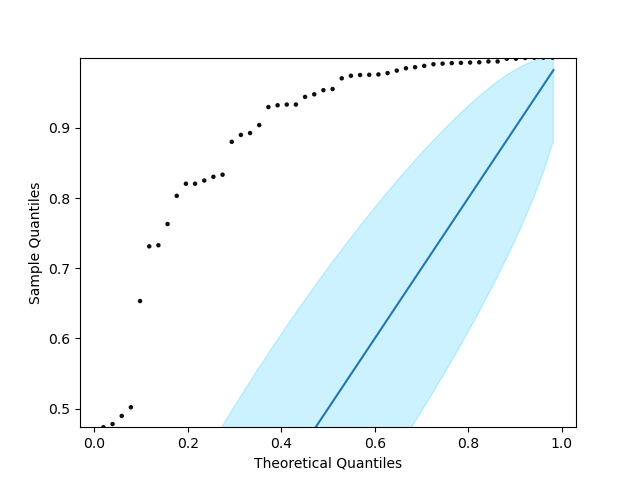}
    \caption{Compensator $\Lambda$ computed using the Linear model \eqref{eq:CompensatorLinear}}
    \label{fig:QQconf_inhib_risktype2}
    \end{subfigure}
    \caption{Test \ref{test:goodness} - Theoretical quantiles of the standard Uniform distribution (blue line) with 95\% confidence intervals (blue region), compared to empirical quantiles of GoF-derived p-values, when simulating a Hawkes Process with parameters $m^\star=1$, $b^\star=2$, and $a^\star=-0.6$, $T_{\max}=5000$ and $n=500$.}
    \label{fig:QQconf_exitation_vs_inhibition}    
\end{figure}
The output of simulations, displayed in Figure \ref{fig:QQconf_exitation_vs_inhibition}, which have been done for $T_{\max}=5000$, shows that the procedure indeed detects that the non-linear model fits accurately the data unlike the linear model. This figure also highlights the importance of the bootstrap procedure in this case, as we can see that the p-values derived from the test of the linear model are stochastically dominated by the uniform law, meaning that taking only one subsample of size $p(n)$ in this case would lead to over accept the linear model.

\subsubsection{Test of the mark assumption}

This section is dedicated to the performance of Tests \ref{Test:OneCoefficient} and \ref{test:goodness} to assess if the mark has an impact on the process. We start by studying Test \ref{Test:OneCoefficient} with the null hypothesis $\Hz:\gamma = 0$. Although this test is very interesting from a practical point of view, as it determines whether a covariate of the model has an impact on the intensity function, it falls outside the theoretical application range of Test \ref{Test:OneCoefficient} as Theorem \ref{Theorem:ConsistanceMLE} has not yet been proven for the marked process. Therefore, we empirically evaluate the procedure by comparing the empirical distribution of the test statistics for this test against a normal distribution when the simulation is done under $\Hz$ and under $\Hu$, the goal being to observe the behavior of probability for both type I and type II risk.

\begin{figure}[!hbtp]
    \centering
    \begin{subfigure}{0.41\textwidth}
    \includegraphics[width =\textwidth]{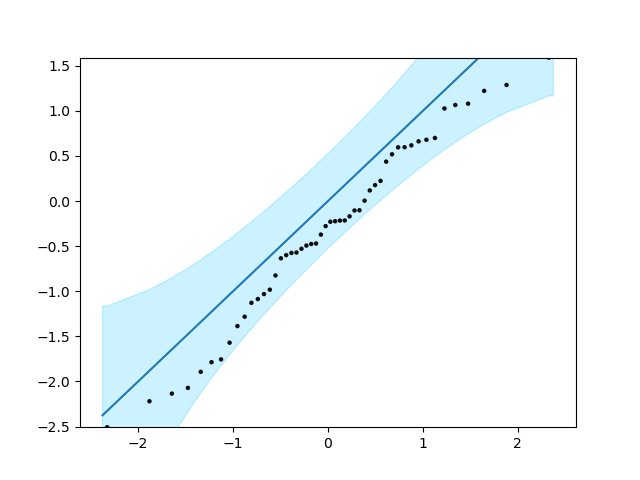}
    \caption{Simulation of a linear process, see Equation \eqref{eq:LMEHP}. }
    \label{fig:Test_Gamma_Null_RiskType1}
    \end{subfigure}
    \quad 
    \begin{subfigure}{0.41\textwidth}
    \includegraphics[width =\textwidth]{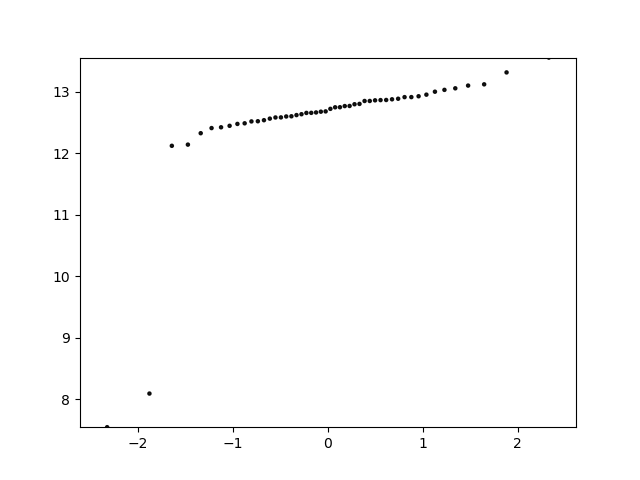}
    \caption{Simulation of a marked linear process, see Equation \eqref{defLamba_mark}. }
    \label{fig:Test_Gamma_Null_RiskType2}
    \end{subfigure}
    \caption{Test \ref{Test:OneCoefficient} - Theoretical quantiles of the standard standard Normal distribution (blue line) with 95\% confidence intervals (blue region), compared to empirical quantiles of $\w{\gamma}/\w{\sigma}$ when simulating either an unmarked Hawkes process (Figure \ref{fig:Test_Gamma_Null_RiskType1}) or a marked linear Hawkes Process (Figure \ref{fig:Test_Gamma_Null_RiskType2}). In both cases, $m^\star=1$, $a^\star=0.6$, $b^\star=1$  $T_{\max}=5000$ and $n=500$ and in the marked scenario $\phigamma(x) = e^{\gammaij x}$ and $(\kappa_i)_i$ i.i.d. following an exponential law of parameter $\gamma^\star = 0.5$.}
    \label{fig:Test_Gamma_Null}    
\end{figure}

Figure \ref{fig:Test_Gamma_Null_RiskType1} shows that, when the process is simulated under $\Hz$, the estimator distribution is close to a normal distribution. This suggests that the asymptotic normality described in Theorem \ref{Theorem:ConsistanceMLE} could be extended in the marked process. 

Conversely, when the parameter $\gamma^*$  is non-zero (meaning when the process is actually a marked process), the mean of $\w{\gamma}$ deviates from zero, and it becomes clear that the test statistics is not a normal law centered on zero, which is what we observe in Figure \ref{fig:Test_Gamma_Null_RiskType2}. 

We then compare those results to the output of the GoF procedure. For this procedure, we conduct simulations in two different frameworks: one where the stationnarity Assumption \ref{Assumption:stationnarity} is taken into account, and another where it is not. 

\begin{figure}[!hbtp]
\centering
\begin{subfigure}{0.4\textwidth}
\includegraphics[width=\textwidth]{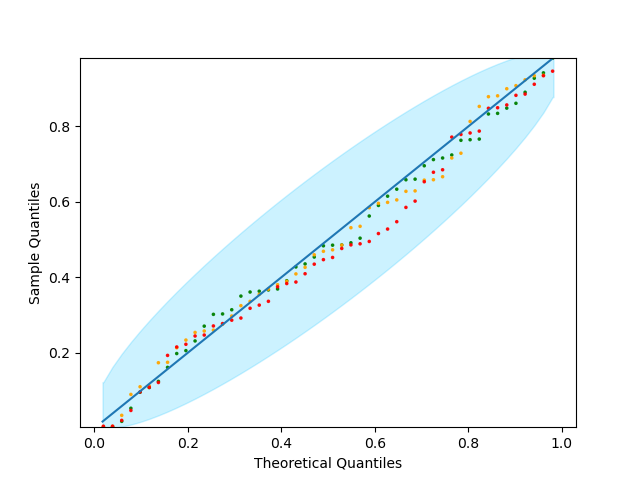}
\caption{Computation is done without using stationary assumption}
\label{fig:GOF_MEHP_non_stationary_ass}
\end{subfigure}
\begin{subfigure}{0.4\textwidth}
\includegraphics[width=\textwidth]{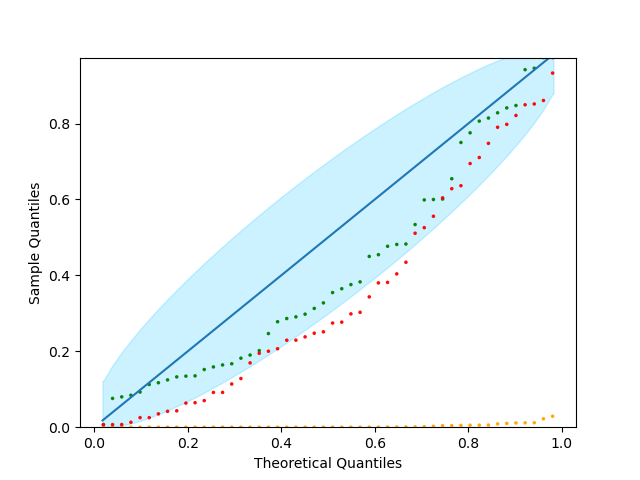}
\caption{Computation is done using stationary assumption}
\label{fig:GOF_MEHP_Stationary_ass}
\end{subfigure}
\caption{Test \ref{test:goodness} - Theoretical quantiles of the standard standard Normal distribution (blue line) with 95\% confidence intervals (blue region), compared to empirical quantiles of GoF-derived p-values when simulating a Marked Hawkes Process with parameters $m^\star=1$, $a^\star=0.6$, $b^\star=1$, $\phi_{\gamma, \psi}(x) = c x^{\gamma}$, and $(\kappa_i)_i$ i.i.d. following an exponential law of parameter $\phi^\star =0.5$, with  $T_{\max}=5000$ and $n=500$. The compensator $\Lambda$ is computed using the Linear model \eqref{eq:CompensatorLinear} without a mark (red points), the Marked Linear model \eqref{eq:CompensatorLinear} with $\phi_{\gamma,\psi}(x) = c x^{\gamma}$ (green points),  the Marked Linear model \eqref{eq:CompensatorLinear} with $\phi_{\gamma,\psi}(x) = c x^{\gamma}$ (orange points). This procedure is performed with and without the stationary assumption. In the case where stationary is not taken into account (Figure \ref{fig:GOF_MEHP_non_stationary_ass}) we have $c=1$, otherwise (Figure \ref{fig:GOF_MEHP_Stationary_ass}) $c$ is the normalisation constant, depending on both $\gamma$ and $\psi$, ensuring Assumption \ref{Assumption:stationnarity}. }
\label{fig:Gof_MEHP}
\end{figure}

Figure \ref{fig:GOF_MEHP_non_stationary_ass} illustrates that when the Assumption \ref{Assumption:stationnarity} is not met, the test fails to distinguish between the different models as in all three scenarios, the hypothesis that the p-values conform to a uniform distribution on $[0,1]$ is accepted. At the opposite, for simulations conducted with the same parameters but incorporating the normalization condition, see Figure \ref{fig:GOF_MEHP_Stationary_ass}, we reject the hypothesis of an unmarked Hawkes process as well as the hypothesis of a Hawkes process with a polynomial parametrization. Given that, for the resampling procedure to be effective, it is imperative for the estimator $\w{\theta}$ of Test \ref{test:goodness} to satisfy $p(n)^{-1/2} \left[ \sum_{i \in S } \Vert \left(\lambda_{\hat{\theta}}\right)_i - (\lambda_\theta)_i \Vert_{l_1([0;T_{\text{max}}])}  \right] \xrightarrow[n \to \infty]{\mathbb{P}} 0$, this suggests that the stationarity assumption is necessary for the marked Hawkes model to ensure that this condition holds.

\begin{figure}[!hbtp]
\centering
\includegraphics[scale = 0.5]{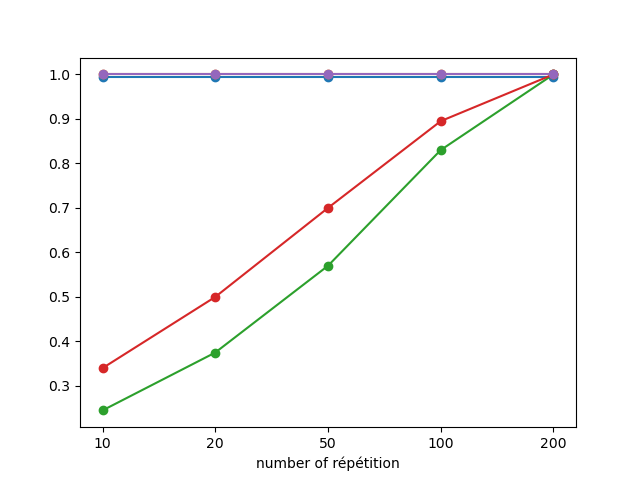}
\caption{Power of the test on the mark when simulating a marked Hawkes process with  $T_{\max}=5000$, $n=500$ repetition and parameters $m^\star=1$, $a^\star=0. 6$, $b^\star=1$, $\phi_{\gamma, \psi}(x) = \frac{\psi - \gamma}{\gamma} e^{x\gamma}$ with $\gamma^\star=0.5$, and $(\kappa_i)_i$ i.i.d. following an exponential distribution with parameter $\psi^\star = 1$. Simulations include a GOF test for Poisson (blue curve), for unmarked exponential Hawkes (green curve), for Hawkes marked with $\phi_{\gamma, \psi}(x) = \frac{\psi - \gamma}{\gamma} e^{x\gamma}$ (red curve) and a bootstrap test (purple curve).}
\label{fig:power_MEHP}
\end{figure}

Figure \ref{fig:power_MEHP} shows the evolution of test power as a function of the number of repetitions in the models (either the number of repetitions available for the GOF test, or the number of bootstrap simulations for the bootstrap test). As expected, power increases with the number of repetitions. However, a notable exception concerns the test associated with the Poisson process and the bootstrap test, whose power is maximal from the first repetition, systematically rejecting the hypothesis of absence of time dependence. This is consistent with the time scale used: as the simulations are carried out in long time, the mark estimator is close to 1, so the bootstrap test always rejects the hypothesis that the mark coefficient is zero, even with few repetitions.

\subsubsection{Test of equality between coefficients.}
The simulations presented here assess the performance of the test $\Hz: ~ b_1^* = b_2^*$ against $\Hu: ~ b_1^* \ne b_2^*$ in the context of a two-dimensional Hawkes Process verifying Assumption \ref{Assumption:Equality_Betaij}. For this simulation, the test statistics is $Z = \frac{( \w{b}_1 - \w{b}_2)}{ \sqrt{\hat{\mathbb{V}}(\w{b}_1) - 2\hat{cov}(\w{b}_1,\w{b}_2) + \hat{\mathbb{V}}(\w{b}_2)} }$.

\begin{figure}[!hbtp]
    \centering
    \begin{subfigure}{0.41\textwidth}
    \includegraphics[width =\textwidth]{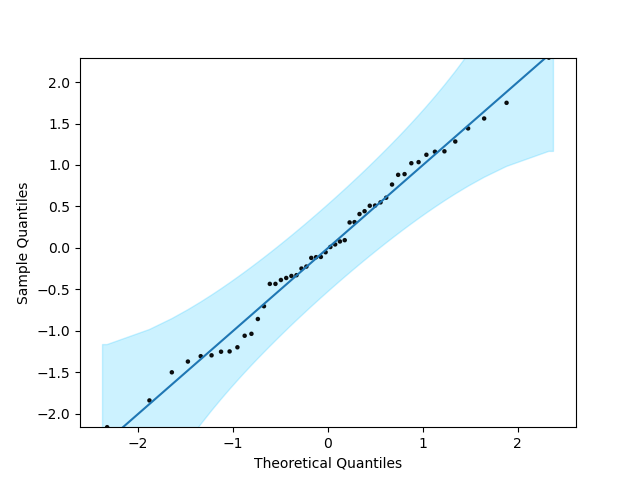}
    \caption{Test of the distribution of the estimator when $b^* = \begin{pmatrix} 1 & 1 \end{pmatrix}^T$ }
    \label{fig:Test_Beta_Equality_RiskType1}
    \end{subfigure}
    \quad 
    \begin{subfigure}{0.41\textwidth}
    \includegraphics[width =\textwidth]{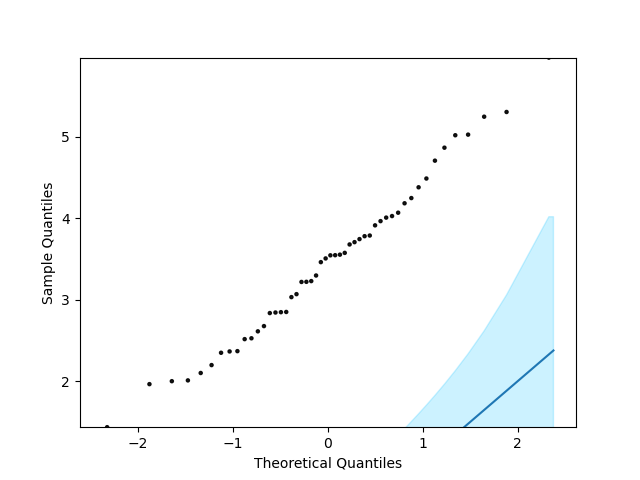}
    \caption{Test of the distribution of the estimator when $b^* = \begin{pmatrix} 1 & 1.5 \end{pmatrix}^T$ }
    \label{fig:Test_Beta_Equality_RiskType2}
    \end{subfigure}
    \caption{Test \ref{test:testdifftheta} - Theoretical quantiles of the standard standard Normal distribution (blue line) with 95\% confidence intervals (blue region), compared to empirical quantiles of $Z$ when simulating a bidimensional Hawkes process with parameter $m^* = \begin{pmatrix} 0.5 & 0.2 \end{pmatrix}^T$, $a^* = \begin{pmatrix} 0.4 & 0.2 \ 0.2 & 0.6 \end{pmatrix}$, and either $b^* = \begin{pmatrix} 1 & 1 \end{pmatrix}^T$ (see Figure \ref{fig:Test_Beta_Equality_RiskType1}) or $b^* = \begin{pmatrix} 1 & 1.5 \end{pmatrix}^T$ (see Figure \ref{fig:Test_Beta_Equality_RiskType2}). In either case,  $T_{\max}=5000$ and $n=500$.}
    \label{fig:Test_Beta_Equality}    
\end{figure}

As seen in Figure \ref{fig:Test_Beta_Equality}, the test statistic indeed follows a centered and standardized normal distribution when the simulation is performed under the null hypothesis. This is consistent with the fact that, when we have simulations derived from a Hawkes model, the estimator indeed follows a normal distribution but centered at $b^*_1 -b^*_2$. Thus, we can easily distinguish cases of equality from other cases by looking into the average of the distribution.

\subsubsection{Computation time}
In this section, we compare the implementation of the test procedures via computation time. We present these results in terms of maximum observation time and, for tests where this is necessary, we generate a number that is the same for all processes and equals to 150.

\begin{figure}[!htbp]
    \centering
    \begin{subfigure}{0.44\textwidth}
         \includegraphics[width=\linewidth]{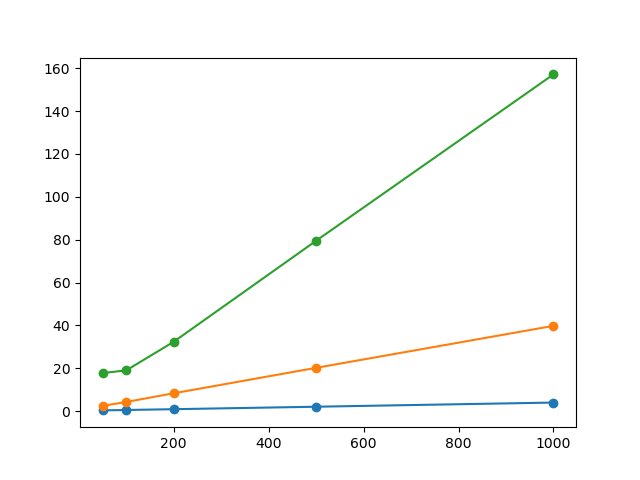}
    \caption{ Test \ref{test:goodness} for a Poisson (blue), a EHP (orange) and a MEHP (green) model.}
    \label{fig:time_computation_GOF}
    \end{subfigure}
    \begin{subfigure}{0.44\textwidth}
    \includegraphics[width = \linewidth]{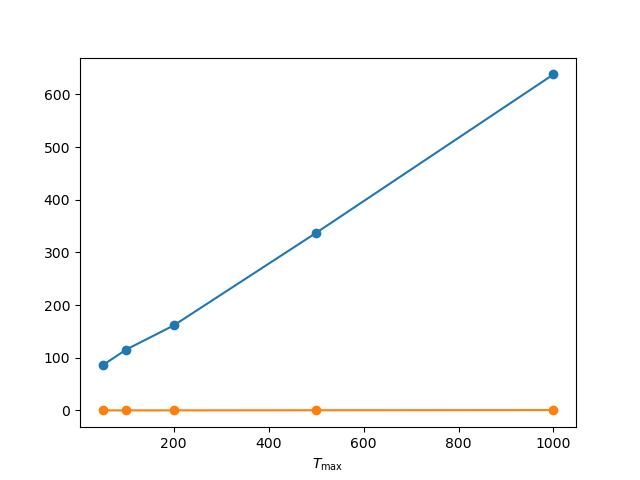}
        \caption{ Test \ref{Test:Boots}  (blue region) and Test \ref{Test:GammaNull} (orange curve)}
    \label{fig:time_computation_zscore_boot}
    \end{subfigure}
    \caption{Time computation (in second) for Test \ref{test:goodness} (Figure \ref{fig:time_computation_GOF}) and both Test \ref{Test:GammaNull} and \ref{Test:Boots} (Figure \ref{fig:time_computation_zscore_boot}) when simulating a MEHP with  $T_{\max}=5000$ and parameters $m^\star=1$, $a^\star=0. 6$, $b^\star=1$, $\phi_{\gamma, \psi}(x) = \frac{\psi - \gamma}{\gamma} e^{x\gamma}$ with $\gamma^\star=0.5$, and $(\kappa_i)_i$ i.i.d. following an exponential distribution with parameter $\psi^\star = 1$. For Test \ref{test:goodness}, $n=150$ repetition are used and for Test \ref{Test:Boots} and $B=150$ repetition are generated.}
    \label{fig:time_computation_Test}
\end{figure}

We begin by analyzing the execution time of Test \ref{test:goodness} for the different models considered: the marked model, the unmarked model and the Poisson model. The most time-consuming step is the estimation phase, which must be repeated for each available repetition of the process. The longer the observation period, the more time-consuming this phase becomes. As shown in Figure \ref{fig:time_computation_GOF}, it is therefore consistent to observe that the more complex the model, the longer the computation time. As shown in Figure \ref{fig:time_computation_zscore_boot}, the computation times associated with Test \ref{Test:Boots} (in Figure \ref{fig:time_computation_zscore_boot}) and \ref{test:goodness} (Figure \ref{fig:time_computation_GOF}) are of the same order of magnitude, especially in comparison of Test \ref{Test:GammaNull} (Figure \ref{fig:time_computation_zscore_boot}). This is because the bootstrap also relies on iterations and requires parameter optimization at for each repetition generated. The main difference, however, lies in the calculation of the likelihood, which is more complex in the bootstrap context. Indeed, to evaluate the likelihood of the $(T_i^b)_i $ times generated, it is necessary to determine to which jump intervals these times belong, a computationally time-consuming operation. This step explains a large part of the difference observed between the execution times of the two tests, the other part being due to the generation of the bootstrap data, which is generally short relative to the optimization time.

\section{Experiments: real-world data}
\label{Section:Realworlddata}

In this section, we present an illustration of the previously presented procedures to publicly available real-world datasets. 

\subsection{Earthquake database}

\begin{figure}[!htpb]
    \centering
        \begin{subfigure}{0.35\textwidth}
        \includegraphics[width = \linewidth]{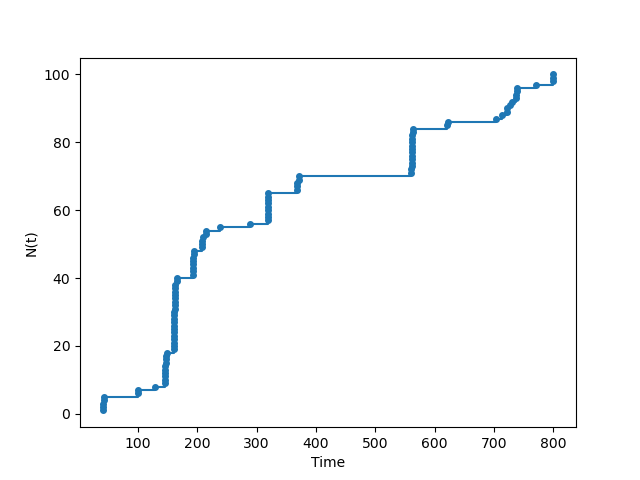}
    \caption{Representation of $t \mapsto N(t)$}
    \label{fig:Nt_ogata}
    \end{subfigure}
    \quad
    \begin{subfigure}{0.35\textwidth}
        \includegraphics[width = \linewidth]{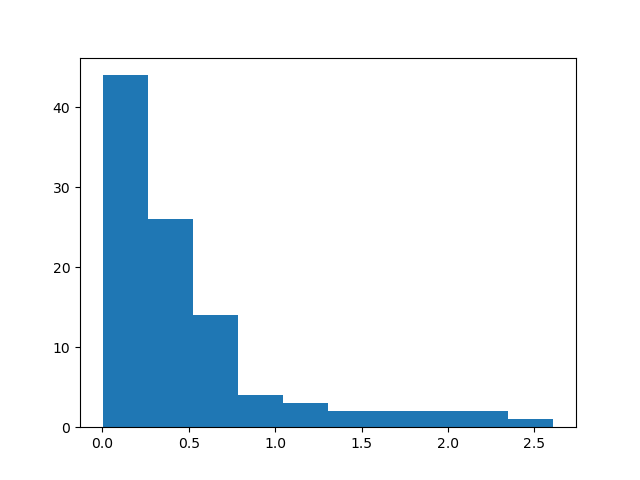}
    \caption{Histogram of the magnitude}
    \label{fig:histo_magnitude}
    \end{subfigure}
    \caption{Data visualization for the \texttt{Ogata} dataset.}
    \label{fig:empirical_representation_ogata}
\end{figure}

We present the results obtained from the analysis of the \texttt{Ogata} dataset, which is part of the \texttt{PTProcess} R package and is commonly used to illustrate inference methods for point processes, in particular Hawkes processes. This example is presented in \cite{harte2010ptprocess}. The dataset contains the occurrence times as well as the magnitudes of a series of earthquakes. The record covers a maximum duration of 800 time units and contains 100 points, each associated with a magnitude. A visualization of the data is provided in Figure \ref{fig:empirical_representation_ogata}, which displays the count process $N(t)$ associated with the occurrences of earthquakes as well as the distribution of the magnitude.

\paragraph{Question of interest and choice of a procedure.}
We aim at investigating whether the magnitude of the previous earthquakes impacts the probability of observing a new occurrence.
For this purpose, we perform both Tests \ref{Test:Boots} and \ref{Test:GammaNull} to assess the mark's impact on the process. It should be noted that Test \ref{Test:OneCoefficient} with $\Hz: \gamma=0$  is not applicable here, since it requires repetitions of the process in order to evaluate the empirical variance of $\w{\gamma}$.

\paragraph{Chosen model.}
Given the empirical density of the magnitude (see Figure \ref{fig:histo_magnitude}), we choose to fit an exponential distribution for the density of the mark: $f_\psi(x) = \psi e^{-\psi x }$.

Besides, the magnitude of an earthquake measures the energy released during the event and is computed from the amplitude of the seismic waves recorded. Since magnitude scales such as Richter or moment magnitude (Mw) are logarithmic, a linear increase in magnitude corresponds to an exponential increase in the energy released. For this reason, we chose to model the impact of the mark on the process by an exponential function, therefore assuming that the influence of an earthquake on the probability of occurrence of future events is proportional to energy released.  As a result, we take  $$\phi_{\gamma, \psi}(x) = \frac{\psi-\gamma}{\psi}e^{\gamma x }.$$

\paragraph{Fitting and test.}
The MLE computation under Model \ref{eq:LMEHP} provides the following estimates $\w{m}=2.065 , \w{a}=11.189 , \w{b}=17.129$, $\w{\gamma}=0.609$ and $ \w{\psi}= 2.065$. First, let us note that the values of both $\w{a}$ and $\w{b}$ are very large suggesting that even though a strong  self-exciting effect is detected, it quickly vanishes, which is consistent with Figure \ref{fig:Nt_ogata}, in which we observe some clusters with many event times but also some time intervals with no event.

Finally, applying  Test\ref{Test:Boots}, we reject the null hypothesis $\Hz : \gamma = 0$, confirming that earthquake magnitude has a significant impact on the process. This result is confirmed by Test \ref{Test:GammaNull}, which gives a p-value of $0.002$.


\subsection{Neuronal activity in mice }

In this section, we focus on the analysis of the data from  \cite{zhang2023cerebellum}, which studies neural activity of mice. In this experiment, several mice explore a circular arena in search of rewards, while their neuronal activity, position, speed and direction are recorded. As each recording consists of five sessions, they are considered as independent repetitions of the process. Therefore, having access to repetitions allows us to apply all the tests presented in Section \ref{Section:Test_Procedure}. We display here the output of Test \ref{Test:OneCoefficient} of the multidimensional model to describe the neural network and perform an equality test on the $(b_i)$ coefficients associated with five neurons, for the fifth recording $140605$ of the mouse $M22$. 



\begin{figure}[!htpb]
    \centering
      \begin{subfigure}{0.4\textwidth}
    \includegraphics[width=\linewidth]{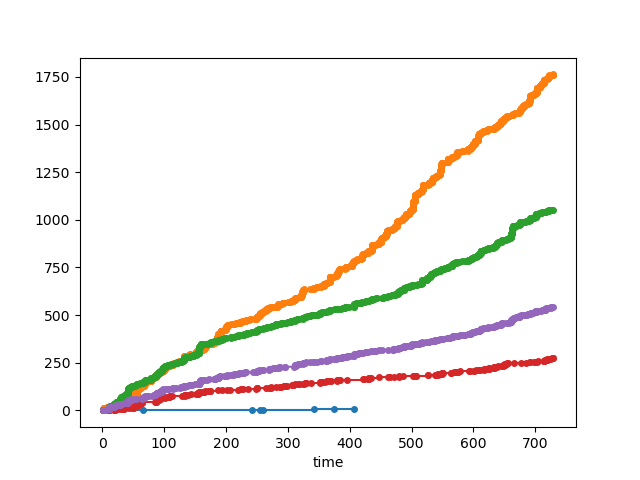}
    \caption{Point process associated to each neuron during the first session.}
    \label{fig:Pt_neuro}
    \end{subfigure}
    \quad
    \begin{subfigure}{0.46\textwidth}
        \includegraphics[width = \linewidth]{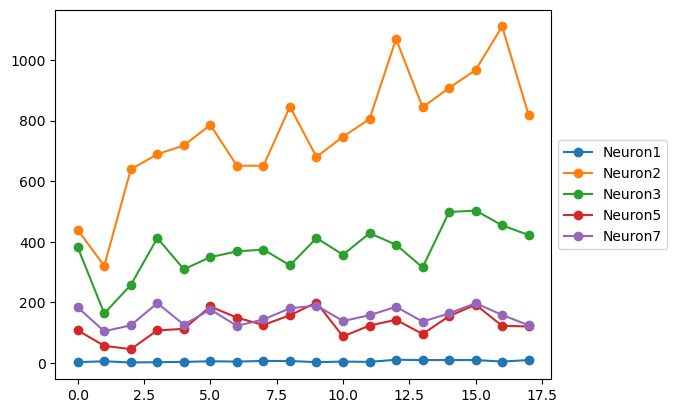}
    \caption{Number of event during each successive period of 200 seconds}
    \label{fig:stationnarity_neuro}
    \end{subfigure}
    \caption{Description of the database.}
    \label{fig:description_neuro}
\end{figure}

The  database contains the activity of five neurons, which is displayed in Figure \ref{fig:description_neuro}. 
As shown in Figure \ref{fig:description_neuro}, the most active neuron (Neuron 2) does not appear to be stationary, as the number of points recorded per 200 seconds indicates a strong increasing trend over time. Because of this non-stationarity, this neuron was excluded from the analysis and thus does not appear in the rest of the analysis.

\begin{table}[!hbtp]
    \centering
    \begin{tabular}{c|c|c|c|c|c|}
        & Neuron 1  & Neuron 3 & Neuron 4  & Neuron 5  \\
        Neuron 1 & NA   &    1e-23 &1e-10 &5e-5\\
       Neuron 3 & & NA     &8e-49&0.92 \\
       Neuron 4 & & & NA   &1e-17\\
       Neuron 5 & & & & NA     \\
    \end{tabular}
    \caption{P-values of the test $b_i = b_j $ for each pair of neurons $i $ and $j $ in the database, for mouse M22, record 140605.}
    \label{tab:pval_neuro_beta_equality}
\end{table}

The table shows the p-values associated with testing $H_0:$ $b_i = b_j$ for each pair of neurons $i \neq j$ in the database, concerning mouse $M22$, record $140605$. The null hypothesis that the $b_i$ coefficients are equal between the different neurons is rejected, suggesting different temporal behavior between them.

\begin{figure}[!hbtp]
    \centering
        \includegraphics[scale =0.45]{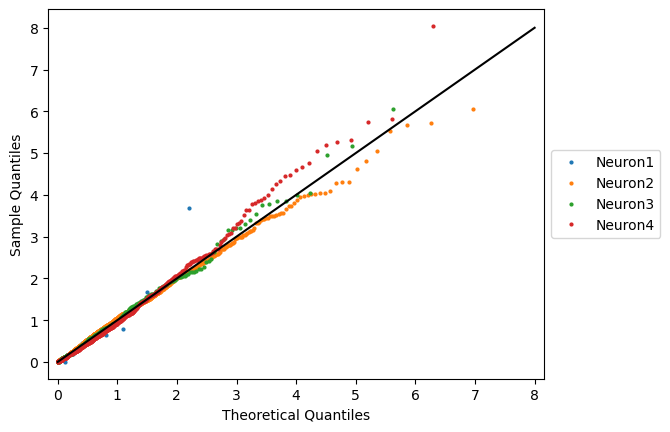}
    \caption{QQ-plot of compensator increments against a unitary exponential distribution for each sub-process considered individually.}
    \label{fig:qqplot_comp_neuro}
\end{figure}

We then perform  Test \ref{test:goodness} in order to assess the global fit between a multivariate Hawkes process and the data: we obtain five p-values with mean value being $10^{-30}$. 
Although we reject the null assumption, Figure \ref{fig:qqplot_comp_neuro} suggests an overall satisfactory fit of the model to the data (it should be remembered, however, that comparing the process increments to those of a compensator via an exponential distribution does not constitute a statistical test in itself see Appendix \ref{appendix:resampling_procedure} and Section \ref{Section:Test_Procedure}). This difference between the two results could also be inherent in the model itself, since, as shown by \cite{bonnet2023inference}, a finer adjustment often requires a procedure for selecting the support for interactions, enabling the interactions between neurons to be identified. 
Another potential explanation for the very low p-value associated with the goodness-of-fit test is the assumption that the memory kernel are exponential. Indeed, neurobiologists have known for a long time that neurons exhibit a refractory period  \citep{refractory1994} that follows a spike, during which a neuron recovers and cannot emit another spike. Modeling the refractory period requires other types of inference procedure, for instance non-parametric estimation methods.

\section{Discussion}\label{Section:discussion}
In this work, we investigate and compare different test procedures based on maximum likelihood approaches for Exponential Hawkes processes. We introduce some extensions to existing theorems related to Marked Hawkes Processes, offering a more robust theoretical framework for their analysis.
 We compile existing test procedures together that we complete in broader frameworks to include nonlinearity and marked versions of the process. We propose indeed a unified resource with a corresponding Python implementation, supported by an extensive numerical study. 
Together with the article, the code is available on a github repository at the address: \url{https://github.com/Msadeler/marked_exp_hawkes}.

One main limitation of this work is that the theoretical properties of the MLE estimator in the nonlinear and marked framework remain unexplored, in particular asymptotic normality properties on which are based many commonly used procedures. If the empirical evidence supports these procedures, developing such theoretical guarantees would complete the present work. Moreover, all procedures described in this article are restricted to the classical exponential kernel and it would be of great interest to develop test procedures in order to determine whether the kernel is exponential or not. 

Furthermore, in this work, we have only investigated small dimensional scenarios even in the multivariate setting, the maximum being a $4$-dimensional Hawkes process for neuronal data. However, as the dimension of the process increases, many challenges may arise. First, the computational cost of some testing methods may become substantially high. In this context, optimizing the likelihood function more efficiently is crucial. For instance, this is the core focus of the recently developed Python library \texttt{Sparklen} (\cite{lacoste2025sparkle}), designed for inference on high-dimensional Hawkes process observations. However, testing procedures are not yet implemented in this library, and this could be an interesting direction for future work.  

Finally, in high-dimensional settings, performing several tests raise the crucial and challenging question of multiple testing. Indeed, if we do not apply appropriate corrections, there is high risk of misinterpreting the outcomes of individual procedures. The most common and simple correction is Bonferroni but is also known to be very conservative. \cite{bonnet2023inference} applied for instance a Benjamini-Hochberg correction procedure to control the false discovery rate \citep{benjamini1995controlling}. Investigating and adapting specifically for point processes such correction methods could also be a valuable avenue for further research.

\paragraph{Acknowledgments.}

This work is part of the 2022 DAE 103 EMERGENCE(S) - PROCECO project supported by Ville de Paris.
The authors thank Adeline Samson for careful reading of the article and Julien Fournier for his help with the neuronal data analysis.

\bibliographystyle{apalike}
\bibliography{biblio.bib}

\begin{appendix} 

\appendixheaderon
\section{Proof of Proposition \ref{Proposition:identifiabilite}}

\label{appendix:proof_identifiability}
We assume that for all $i \in \{1,...,d\}$, for all $j \neq i$, there exist $k,l >0 $ such that $T_{k}$ and $T_{\ell}$ are arrival times of component $j$ such that $\kappa_k \neq \kappa_l$.
Let us consider the link function $\phi_{\gammaij }:x \mapsto e^{\gammaij  x}$. Note that the case $\phigamma(x) = x^{\gammaij }$ can be treated in the same manner.

Suppose that there exist $ \theta_i = ( m_i, a_{ij}, \bij, \gammaij )$ and $ \widetilde{\theta}_{i} = ( \widetilde{m}_i, \widetilde{a}_{ij}, \widetilde{b}_{ij}, \widetilde{\gamma }_{ij} )$ such that $\lambda_{i,\theta} = \lambda_{ i,\widetilde{\theta}}$. Then, $  m_i =  \lambdatheta{i} (T_{1}) = \lambda_{ \widetilde{\theta}_i}(T_{1}) = \widetilde{m}_i$, hence $ \widetilde{m}_i =  m_i$.

Furthermore, for all $t \in (T_{1},T_{2})$, $$\lambda_{i,\theta}(t) = m_i + a_{im_1} \phi_{\gamma_{im_1}}\left(\kappa_1\right) e^{-b_{im_l}(t-T_1)}$$ where $m_1 \in \{1, \ldots, d\}$ is the component from which the event $T_1$ comes, and, likewise, 
$$\lambda_{i,\widetilde{\theta}}(t) = \widetilde{m_i} + \widetilde{a}_{im_1} \phi_{\widetilde{\gamma}_{im_1}}\left( \kappa_1\right) e^{-\widetilde{b}_{im_1}(t-T_1)} .$$
Thus, as $m_i = \widetilde{m}_i$, we obtain that $\widetilde{ a}_{im_1}\phi_{\widetilde{\gamma}_{im_1} } ( \kappa_1) = a_{im_1} \phi_{\gamma_{im_1} } ( \kappa_1)$ and  $\widetilde{b}_{im_1} = b_{im_1}.$

Considering that, for all $k > 1$ and for all $t \in \left( T_k, T_{k+1} \right)$, $$\lambda_{i,\theta}(t) = m_i +  \sum_{\ell=1}^{k}a_{im_\ell} \phi_{\gamma_{im_\ell}}\left( \kappa_\ell\right) e^{-b_{im_\ell}(t-T_\ell)}$$ (and similarly for $\lambda_{\widetilde{\theta}_i}$), we can use a strong recurrence to show that for all $\ell$:
$$\widetilde{b}_{im_\ell} = b_{im_\ell} ~ \text{and} ~ a_{im_l} \phi_{\gamma_{im_\ell}}(\kappa_\ell) =  \widetilde{ a}_{im_\ell} \phi_{\widetilde{\gamma}_{im_\ell} } ( \kappa_\ell).$$

Let $(i ,j)\in \{1,...,d\}^2$, such that $i \neq j$, and let $T_{k},T_{\ell}$ be two arrival times associated with component $j$ (meaning $m_\ell = m_\ell = j$) such that $\kappa_k \neq \kappa_\ell$.\\

Let us take $t \in (T_{k},T_{k+1})$. 
The intensity of the process at time $t$ writes,
\begin{align*}
        \lambdatheta{i,\star} (t) &=  m_i + \sum_{\ell=1}^k a_{im_{\ell}} \phi_{\gamma_{im_{\ell}}}(\kappa_l) e^{-b_{i} ( t-T_{\ell})} \\
        & = m_i + e^{-b_i( t-T_{k})} \sum_{\ell=1}^k a_{im_{\ell}} \phi_{\gamma_{im_{\ell}}}(\kappa_\ell) e^{-b_{i} ( T_{k}-T_{\ell})}  \\
        & = m_i +  e^{-b_i( t-T_{k})} \left(  a_{im_{k}}\phi_{\gamma_{im_{k}}}(\kappa_k) + \sum_{l=1}^{k-1} a_{im_{\ell}} \phi_{\gamma_{im_{\ell}}}(\kappa_\ell) e^{-b_{i} ( T_{k}-T_{\ell})} \right) \\
        & = m_i +  e^{-b_i( t-T_{k})} \left(  a_{im_{k}}\phi_{\gamma_{im_{k}}}(\kappa_k) + \lambdatheta{i}( T_{k}) -m_i \right) \\
        & = m_i +  e^{-b_i( t-T_{k})}\left( \lambdatheta{i,\star}( T_{k}^{+}) - m_i \right).
    \end{align*}

Then, as the same type of equality holds for $\lambda_{i,\widetilde{\theta}}$, we have $$ \widetilde{ a}_{ij} \phi_{\widetilde{\gamma}_{ij}} ( \kappa_k) = a_{ij} \phi_{\gamma_{ij}} ( \kappa_k) .$$
And similarly at $T_{\ell}$:
$$\widetilde{ a}_{ij} \phi_{\widetilde{\gamma }_{ij}} ( \kappa_\ell) = a_{ij} \phi_{\gammaij } ( \kappa_\ell).$$

Let us recall that $\phigamma$ either takes the form of $x \mapsto e^{\gammaij x}$ or $x \mapsto x^\gammaij$ and is therefor submultiplicative with respect to $\gammaij$.

Therefore, $\widetilde{a}_{ij} \phi_{\widetilde{\gamma}_{ij }- \gammaij } ( \kappa_k)=   a_{ij} $ and $\widetilde{a}_{ij}\phi_{\widetilde{\gamma}_{ij }- \gammaij } ( \kappa_\ell)=   \aij $. As $\widetilde{a}_{ij}>0$, we obtain $\phi_{\widetilde{\gamma}_{ij }- \gammaij } ( \kappa_\ell) = \phi_{\widetilde{\gamma}_{ij }- \gammaij } ( \kappa_k)$.

Finally, considering the parametric shape $\phi_{\gammaij}$ if $\phi_{\gammaij}(\kappa_1) = \phi_{\gammaij}(\kappa_2) $ with $\kappa_1,\kappa_1>0$ and $\kappa_1 \neq \kappa_2$, then $\gamma=0$. Hence, $\widetilde{\gamma}_{ij}- \gammaij  = 0 $, i.e., $\widetilde{\gamma}_{ij} = \gammaij $. 

Consequently, $a_{ij} = \widetilde{a}_{ij}$. Thus, the model is identifiable.

\appendixheaderon

\appendixheaderon
\section{Proof of Theorem \ref{Theorem:TimeChangeMark}}
\label{appendix:proof_time_change}
The proof presented here closely follows the demonstration of Theorem 14.6.4 in \cite{daley2008introduction}.

\noindent
We will use the fact that the probability generating function of our process determines its distribution and show that the probability generating function of the process $\overline{N}$ is actually equal to the probability generating function of a Poisson process on $\mathbb{R}_+ \times [0,1]$.

\noindent
Let $h$ be a continuous function from $\mathbb{R_+} \times \mathbb{R}$ to $[0,1]$ such that $1-h$ has bounded support.

\noindent
We aim to show that 
\[ 
\left[ 
 \prod_{i=1}^{+\infty}  h(\Lambda_g(t_i), F(\kappa_i))  
\right]  = \int_{\mathbb{R}_+^* \times (0,1)} [h(t, \kappa) -1 ] dt d\kappa.
\]

\noindent
We define: 
\begin{align*}
H^\mathcal{F}(t) = &  \left[ 
 \prod_{1 \leq T_i \leq t }  h(\Lambda_g(t_i), F(\kappa_i))  
\right]  e^{  - \int_{(0,t) \times \mathbb{R}} [h( \lambda_g(s), F(\kappa)) -1 ] \lambda(t, \kappa) dk ds},
 \\
 u^{\mathcal{F}}(s) = &  \int_{\mathbb{R}} [h(\Lambda_g(s),F(\kappa)) -1 ]f(\kappa \vert s)  dk = \overline{h}(s) -1.
\end{align*}

We define the process $G$ by 
\[ 
G^\mathcal{F}(t_i) - G^\mathcal{F}(t_i^-) =  h(\Lambda_g(t_i), F(\kappa_i)) / \overline{h}(t_i) \text{ and } G^\mathcal{F}_c(s) =  - \lambda_g(s) ds \text{ for } s \notin \{t_i\},
\]
with  
\[ 
G^{\mathcal{F}}_{c}(t) = G^{\mathcal{F}}(t) - \sum_{t_i \leq t} \left(G^\mathcal{F}(t_i) - dG_{c}^{\mathcal{F}} (t_i^-) \right) 
\]
the continuous part of $G$.

According to \cite{daley2003introduction} (page 107, Lemma 4.6.2), since $G$ is right-continuous, increasing, and $u^\mathcal{F}$ is measurable, then $H^\mathcal{F}$ is the unique solution of the equation:
\begin{equation}
\label{H^F}
    H^\mathcal{F}(t)= 1 + \int_{(0,t)} H^\mathcal{F}(s-) u^\mathcal{F}(s)dG^\mathcal{F}(s),
\end{equation}
(because $\sup_{0 \leq s \leq t} |H(s)| \leq \left(|| h-1 ||_\infty + 1 \right)^{N(t)} \times e^{||h-1||_\infty \int_{(0,t) \times \mathbb{R}} \lambda_g(s,\kappa) ds d\kappa} < \infty$).

\noindent
Now, we have 
\[ 
dG^\mathcal{F} (s)  = \int_\mathbb{R} \left( \frac{h( \Lambda_g(s) , F(\kappa) )}{\overline{h}(s) }N(ds \times d\kappa) - \lambda_g(s)\right) ds,
\]
so $G^\mathcal{F}$ is a martingale. Let $\Phi(h) = \lim_{t \longrightarrow \infty} H^\mathcal{F}(t)$.

\noindent
Taking the limit in \ref{H^F} as $t$ tends to infinity, and then the expectation, we obtain the following result:
\begin{align}
\label{limH^F}
    \mathbb{E}( \phi( h) ) = 1 + \mathbb{E} \left[ \int_{\mathbb{R}^+} H^\mathcal{F}(t-)[\overline{h}(t) -1] dG^\mathcal{F}(t) \right].
\end{align}

\noindent
Now:
\begin{align*}
\mathbb{E} \left[ \int_{\mathbb{R}^+} H^\mathcal{F}(t-)\left[\overline{h}(t) -1\right] dG^\mathcal{F}(t) \right] & = \mathbb{E} \left[ \int_{\mathbb{R}^+} H^\mathcal{F}(t-) \int_\mathbb{R} [ h( \lambda_g(t), \kappa) -1] f(\kappa \vert t ) d\kappa \text{ }dG^\mathcal{F}(t) \right]
\\ 
& = \int_{\mathbb{R}_+} \mathbb{E} \left[ \int_{\mathbb{R}_+} H^\mathcal{F}(t-) \left[ h(\Lambda_g(t), F(\kappa))-1 \right] f(\kappa \vert t ) dG^\mathcal{F}(t) \right] d\kappa
\\
& = 0,
\end{align*}
where the last equality holds because $G^\mathcal{F}$ is a martingale.

\noindent
Finally, equation \ref{limH^F} becomes:
\begin{align}
\label{Eq2}
    1 = \mathbb{E} \left[\prod_{t_i} h( \Lambda_g (t_i) , F(\kappa_i) ) e^{- \int_{(0,\infty) \times \mathbb{R}} [h( \Lambda_g(s), F(\kappa)) -1 ] \lambda(t, \kappa) dk ds } \right].
\end{align}

\noindent
We show that 
\[ 
\int_{(0,\infty) \times \mathbb{R}} [h( \Lambda_g(s), F(\kappa)) -1 ] \lambda_g(t) f(\kappa \vert t) dk ds = \int_{(0,\infty) \times [0,1]} [h( s, \kappa) -1 ] ds d\kappa.
\]

\noindent
We know that $\kappa \mapsto F(\kappa \vert t)$ is $\mathcal{C}^0$ for all $t$.

\noindent
Therefore, if $X$ has density $\kappa \mapsto f(\kappa \vert t)$, $F(X \vert t) \sim \mathcal{U}([0,1])$.

\noindent
Thus, by the transfer theorem:
\[ 
\int_{\mathbb{R}} [h(\Lambda_g(t), F(\kappa \vert t))-1] f(\kappa \vert t) d\kappa = \int_{(0,1)} [h(\Lambda_g(t),\kappa)-1] d\kappa.
\]

By Fubini's theorem, we have:
\begin{align*}
    \int_{(0,\infty) \times \mathbb{R}} [h( \Lambda_g(s), F(\kappa))-1] \lambda_g(t) f(\kappa \vert t) dk ds & = \int_{(0,\infty) \times (0,1)} [h( \Lambda_g(s),\kappa)-1] \lambda_g(t) d\kappa ds
    \\
    & = \int_{(0,1)} \left(\int_{\mathbb{R_+}} [h( \Lambda_g(s),\kappa)-1] \lambda_g(t) dt \right) d\kappa
    \\
    & = \int_{(0,1)} \left(\int_{\phi(\mathbb{R_+})} [h( s,\kappa)-1] (\phi_* \mu)(ds) \right),
\end{align*}
with $\phi(x) = \Lambda_g(x)$ and $\mu(A) = \int_A \lambda_g(t) dt = \Lambda_g(A)$. Thus, $\phi_* \mu = \text{Leb}_{\vert \Lambda_g(\mathbb{R_+})}$.

\noindent
Therefore:
\begin{align*}
     \int_{(0,\infty) \times \mathbb{R}} [h( \Lambda_g(s), F(\kappa))-1] \lambda_g(t) f(\kappa \vert t) dk ds & = \int_{(0,1)} \int_{\Lambda_g(\mathbb{R_+})} [h(s,\kappa)-1] ds d\kappa.
\end{align*}

\noindent
Thus, since $\lim_{t \rightarrow +\infty}\Lambda_g(t) \rightarrow +\infty$ and $\Lambda_g(0)=0$, and $t \mapsto \Lambda_g(t)$ is $C^0$, we have:
\begin{align*}
     \int_{(0,\infty) \times \mathbb{R}} [h( \Lambda_g(s), F(\kappa))-1] \lambda_g(t) f(\kappa \vert t) dk ds & = \int_{(0,1)} \int_{\mathbb{R_+}} [h(s,\kappa)-1] ds d\kappa.
\end{align*}

So finally, returning to \ref{Eq2}, since the calculated integral above is no longer random, we can take it out of the expectation and obtain:
\[ 
\mathbb{E} \left[\prod_{t_i} h(\Lambda_g(t_i), F(\kappa_i \vert t_i)) \right] = \exp \left( \int_{ \mathbb{R}_+ \times [0,1]}[h(t, \kappa) - 1] ds d\kappa \right).
\]
This completes the proof.

\appendixheaderon
\section{Proof of proposition \ref{Proposition:CompensateurNonLinear}}
    \label{Demo:ComposateurMultiMark}

Let us start by showing that $T^{i,*}_{k} = \min \left(T_{k}  + b_i^{-1} \log \left( \frac{m_i - \lambdatheta{i,\star}(T_{k}^+) }{m_i} \right) \ensuremath{\mathds{1}}_{ \lambdatheta{i,\star}(T_{k}^+) <0}, T_{k+1} \right) $.
Let us recall (see Appendix \ref{appendix:proof_identifiability}) that, for $t \in (T_k , T_{k+1})$, $\lambda_{\theta_{i}}^* (t) = m_i +  e^{-b_i( t-T_{k})}\left( \lambdatheta{i,\star}( T_{k}^{+}) - m_i \right)$.

On one hand, if $\lambdatheta{i,\star}( T_{k}^+) \geq 0  $, then by definition $ T^{i,*}_{k}=T_{k} $ 
As a result, 
$$ T^{i,*}_{k} = \min \left(T_{k}  + b_i^{-1} \log \left( \frac{m_i - \lambdatheta{i,\star}(T_{k}^+) }{m_i} \right) \ensuremath{\mathds{1}}_{ \lambdatheta{i,\star}(T_{k}^+) <0}, T_{k+1} \right).$$
On the other hand, if $\lambdatheta{i,\star}( T_{k}^+) \leq 0  $, then  $\lambdatheta{i,\star}( T_{k}^+)  - m_i <0$ and $\lambdatheta{i,\star}$ is increasing on $(T_{k},T_{k+1})$. 
Therefore, if there is a point for which $\lambdatheta{i,\star}$ is positive, this point $t^*$ verifies:
$$m_i +  e^{-b_i( t^*-T_{k})}\left( \lambdatheta{i,\star}( T_{k}^+) -m_i \right) = 0 \iff t^* = T_{k}  + b_i^{-1} \log \left( \frac{m_i - \lambdatheta{i,\star}(T_{k}^+) }{m_i} \right) \ensuremath{\mathds{1}}_{ \lambdatheta{i,\star}(T_{k}^+) <0}.$$
Finally, we obtain that 
$$T^{i,*}_{k} = \min \left(T_{k}  + b_i^{-1} \log \left( \frac{m_i - \lambdatheta{i,\star}(T_{k}^+) }{m_i} \right) \ensuremath{\mathds{1}}_{ \lambdatheta{i,\star}(T_{k}^+) <0}, T_{k+1} \right). $$

Let us now explicit $\Lambdatheta{i}(T)$.
We know that 
$$\Lambdatheta{i}(T) = \int_{(0,T_{1})} \lambdatheta{i}(t) dt + \sum_{ k= 1}^{N(T)} \int_{(T_{k},T_{k+1})} \lambdatheta{i}(t) dt + \int_{T_{N(T)}}^T \lambdatheta{i}(t) dt .$$
And, $\lambdatheta{i}(t) = \max \left( \lambdatheta{i,\star}(t), 0 \right)$ and $\lambdatheta{i}(t) = \lambdatheta{i,\star}(t) \iff  t \in ( T^{i,*}_{k}, T_{k+1}) $ and $\lambdatheta{i}(t) = 0$ otherwise.
As a result, 
$$\Lambdatheta{i,\star}(T) = \int_{(0,T_{1})} \lambdatheta{i,\star}(t) dt + \sum_{ k= 1}^{N(T)} \int_{(T^{i,*}_{k},T_{k+1})} \lambdatheta{i,\star}(t) dt + \int_{T^{i,*}_{N(T)}}^T \lambdatheta{i,\star}(t) dt.$$
Using that
$ \int_{(0,T_{1})} \lambdatheta{i,\star}(t) dt = m_1 T_{1}$ and the previous computation, we obtain the following equality. 
\begin{align*}
    \int_{T^{i,*}_{k}}^{T_{k+1}} \lambdatheta{i,\star}(t) dt  & = \int_{(T^{i,*}_{k},T_{k+1})} m_i +  e^{-b_i( t-T_{k})}\left( \lambdatheta{i}( T_{k}^+) -m_i \right) dt \\
    & = m_i( T_{k+1} - T^{i,*}_{k}) +  b_i^{-1} \left( \lambdatheta{i}(T_{k}^+)- m_i \right)\left[ e^{ -b_i (T_{k}^{i,*}- T_{k})} - e^{ -b_i \left(  T_{k+1}-T_{k} \right)} \right]   \\
    & = J_k.
\end{align*} 
And, with the same idea:
\begin{align*}
    \int_{T^{i,*}_{N(T)}}^T \lambdatheta{i,\star}(t) dt & = m_i \left[  T -  T_{N(T)}^{i,*} \right] + b_i^{-1} \left( \lambdatheta{i}(T_{N(T)}^+)- m_i \right)\left[ e^{ -b_i (T_{N(T)}^{i,*}- T_{N(T)})} - e^{ -b_i \left(T -T_{N(T)} \right)} \right]\\
     & = J_{N(T)}.
\end{align*} 
As a consequence, we have shown that:  
\[ \lambdatheta{i}(T) = \begin{cases} 
          m_i T & \text{si } T \leq T_{1} \\
          m_i T_{1} +  \left(\sum_{k=1}^{N(t)} J_k\right)  & \text{ otherwise }.
       \end{cases} 
    \]

\appendixheaderon
\section{Illustration of the resampling procedure}
\label{appendix:resampling_procedure}

\begin{figure}[!hbtp]
\centering
        \begin{subfigure}{0.31\textwidth}
        \centering
        \includegraphics[width =\textwidth]{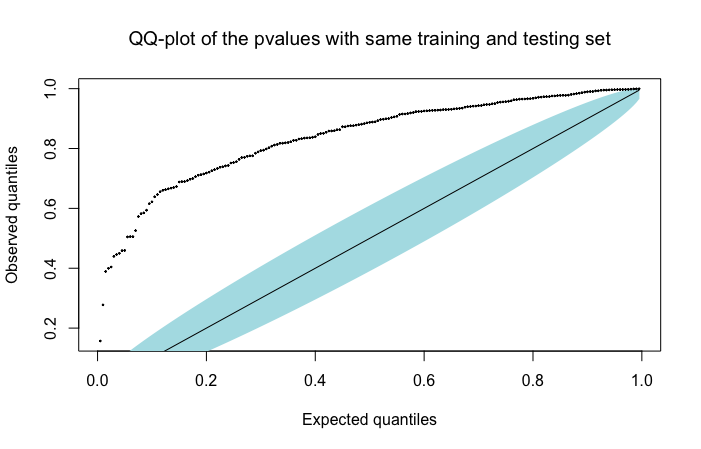}
        \caption{Same sample}
        \label{fig:GoF_same_sample}
    \end{subfigure}
    \quad
    \begin{subfigure}{0.31\textwidth}
        \centering 
        \includegraphics[width =\textwidth]{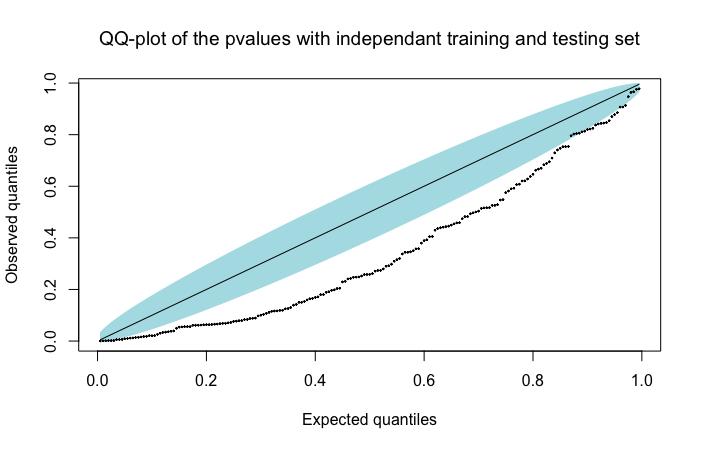}
        \caption{Independent samples}
        \label{fig:GoF_inde_sample}
    \end{subfigure}
\quad
    \begin{subfigure}{0.31\textwidth}
        \centering 
        \includegraphics[width =\textwidth]{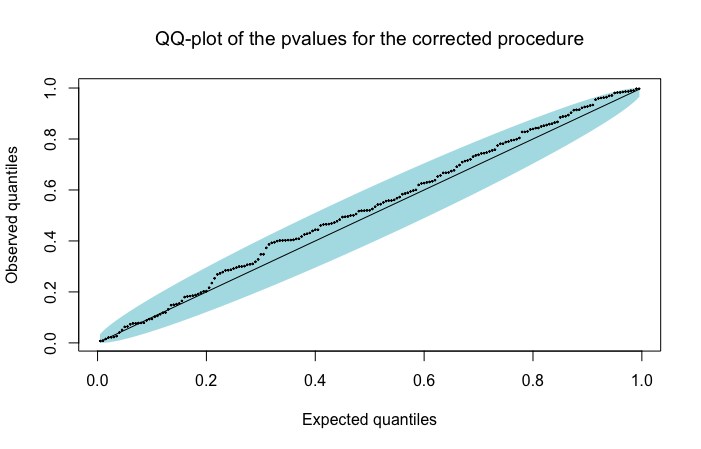}
        \caption{Resampling procedure}
        \label{fig:GoF_resample_procedure}
    \end{subfigure}
        \caption{Uniformity Test of GoF-derived p-values, using a qqconf-plot, when simulating a marked Hawkes Process with $m^\star=1$, $a^\star=0.6$, $b^\star=1$ and estimating on the same sample (Figure \ref{fig:GoF_same_sample}), on an independent sample (Figure \ref{fig:GoF_inde_sample}) or using the corrected Test \ref{test:goodness} (Figure \ref{fig:GoF_resample_procedure}).}
        \label{fig:GoF_procedures}
\end{figure}

We display here the issues associated with the conventional practice of the goodness-of-fit test that appear in situations where the time change theorem is applied either on the same sample or on an independent sample than the one used to compute the MLE. To that end, we present in Figure \ref{fig:GoF_procedures}, via a qqconf plot, the results of the tests verifying that the distributions of p-values obtained follow a uniform distribution.

As demonstrated by Figures \ref{fig:GoF_same_sample} and \ref{fig:GoF_inde_sample}, the conventional application of the GoF test often results in either the over-acceptance or under-acceptance of the null hypothesis, depending on whether the test is conducted on the same sample or an independent sample. Test \ref{test:goodness} allows to address this issue using a subsampling and a bootstrap procedure, as we notice that p-values obtained thought this test follow uniform distribution Figure \ref{fig:GoF_resample_procedure}. Consequently, for the remaining simulations, we exclusively used this procedure when referring to the GoF test.

\end{appendix}

\end{document}